\documentclass[fleqn,usenatbib]{mnras}

\usepackage{newtxtext,newtxmath}
\usepackage{wasysym}
\usepackage{soul}
\usepackage{xcolor}
\usepackage[T1]{fontenc}

\DeclareRobustCommand{\VAN}[3]{#2}
\let\VANthebibliography\thebibliography
\def\thebibliography{\DeclareRobustCommand{\VAN}[3]{##3}\VANthebibliography}

\usepackage{graphicx}	% Including figure files
\usepackage{amsmath}	% Advanced maths commands
\title[J. Dou et al.]{Formation of super-Mercuries via giant impacts}

\author[J. Dou et al.]{
Jingyao Dou,$^{1}$\thanks{E-mail: jingyao.dou@bristol.ac.uk}
Philip J. Carter$^{1}$, and
Zoë M. Leinhardt$^{1}$
\\
$^{1}$School of Physics, H.H. Wills Physics Laboratory, University of Bristol, Bristol BS8 1TL, UK \\
}

\date{Accepted XXX. Received YYY; in original form ZZZ}

\pubyear{2024}

\begin{document}
\label{firstpage}
\pagerange{\pageref{firstpage}--\pageref{lastpage}}
\maketitle

\begin{abstract}
During the final stage of planetary formation, different formation pathways of planetary embryos could significantly influence the observed variations in planetary densities. Of the approximately 5,000 exoplanets identified to date, a notable subset exhibit 
core fractions reminiscent of Mercury, potentially a consequence of high-velocity giant impacts. In order to better understand the influence of such collisions on planetary formation and compositional evolution, we conducted an extensive set of smoothed particle hydrodynamics giant impact simulations between two-layered rocky bodies. These simulations spanned a broad range of impact velocities from one to eleven times the mutual escape velocity. We derived novel scaling laws that estimate the mass and core mass fraction of the largest post-collision remnants. Our findings indicate that the extent of core vaporization markedly influences mantle stripping efficiency at low impact angles. We delineate the distinct roles played by two mechanisms -- kinetic momentum transfer and vaporization-induced ejection -- in mantle stripping. 
Our research suggests that collisional outcomes for multi-layered planets are more complex than those for undifferentiated planetesimal impacts. Thus, a single universal law may not encompass all collision processes. We found a significant decrease in the mantle stripping efficiency as the impact angle increases. To form a 5 M$_{\oplus}$ super-Mercury at $45^{\circ}$, an impact velocity over 200 km s$^{-1}$ is required. This poses a challenge to the formation of super-Mercuries through a single giant impact, implying that their formation would either favor relatively low-angle single impacts or multiple collisions.

\end{abstract}

\begin{keywords}
planets and satellites: formation -- hydrodynamics -- planets and satellites: composition -- planets and satellites: terrestrial planets -- methods: numerical
\end{keywords}

%%%%%%%%%%%%%%%%%%%%%%%%%%%%%%%%%%%%%%%%%%%%%%%%%%

%%%%%%%%%%%%%%%%% BODY OF PAPER %%%%%%%%%%%%%%%%%%

\section{Introduction}

Several of rocky exoplanets discovered to date, including Kepler-107\,c \citep{bonomo_giant_2019}, GJ~367\,b \citep{lam_gj_2021}, Kepler-406\,b \citep{marcy_masses_2014}, K2-38\,b \citep{toledo-padron_characterization_2020}, and K2-229\,b \citep{santerne_earth-sized_2018}, have very high bulk densities, suggesting that they are iron-rich and may have a composition similar to that of Mercury. These planets are often referred to as "super-Mercuries" due to their measured mass being at least one order of magnitude larger than Mercury's. The formation mechanisms of Mercury and super-Mercuries are still debated and could be the result of mantle photo-evaporation \citep{cameron_partial_1985,fegley_vaporization_1987}, the separation of silicate and iron particles in the initial inner disk \citep{weidenschilling_iron_1978, wurm_photophoretic_2013, johansen_nucleation_2022-1}, or giant impacts \citep{benz_collisional_1988,benz_origin_2007,chau_forming_2018,bonomo_giant_2019,reinhardt_forming_2022}. According to \citet{ito_hydrodynamic_2021}, photo-evaporation of mantle material is not significant enough to greatly affect the bulk composition of a 1 M$_{\oplus}$ rocky planet. Although photophoretic forced separation could explain the formation of Mercury, it cannot accurately model Kepler-107 c, which is more than twice as dense (about 12.6 g cm$^{-3}$) as the innermost Kepler-107 b (about 5.3 g cm$^{-3}$). Instead, the dissimilar densities are consistent with a giant impact event that resulted in striping the silicate mantle off of the precursor planet resulting in a dense Kepler-107 c \citep{bonomo_giant_2019}.

Giant impacts (violent collisions between planet-sized bodies) are extremely energetic events that could strip away a large fraction of mantle and generate a post-collision body with a high core fraction. \citet{benz_collisional_1988,benz_origin_2007,asphaug_mercury_2014,chau_forming_2018} previously examined several impact conditions that could have contributed to the formation of Mercury. Mercury's current core fraction could have been achieved by a single giant impact, in which the proto-Mercury was hit by a smaller projectile \citep{benz_origin_2007}, or by a massive one like Venus or Earth after which Mercury was the left remnant from the impactor \citep{asphaug_mercury_2014}, or by a sequence of less energetic giant impacts \citep{chau_forming_2018}. 
The newly discovered super-Mercury exoplanets exhibit a wide range of mass and core fractions, and little is known about their formation history. There is currently no reliable way to link an observed planet with the specific conditions of the impact that led to its formation. 

Scaling laws provide a means to predict collision outcomes based on the impact parameters, and can thus reveal which kind of collisions can produce high-density planets. \citet{stewart_velocity-dependent_2009} derived a universal law for the post-collision mass of the largest remnant for planetesimal-planetesimal collisions up to radii of 50 km. Based on about 60 smoothed particle hydrodynamics (SPH) simulations of giant impacts with targets ranging from 1 M$_{\oplus}$ to 10 M$_{\oplus}$, \citet{marcus_collisional_2009} found that the universal law was also in agreement with head-on impacts in the super-Earth regime. Additionally, they derived a scaling law to predict the iron mass fraction of the post-collision remnant for head-on impacts. Following this work, \citet{leinhardt_collisions_2012} made one of the most comprehensive and widely used scaling laws that can predict the outcomes of planetesimal impacts with different impact angles and impactor to target mass ratios. However, the data for collisions between planet-size bodies were adopted from several previous studies that had different target composition setups. Therefore, \citet{leinhardt_collisions_2012} did not provide scaling laws to predict the iron mass fraction of post-collision remnants. In contrast, \citet{carter_collisional_2018} simulated impacts between differentiated targets of sizes that fall between the size regimes explored by \citet{marcus_collisional_2009} and \citet{leinhardt_collisions_2012}. \citet{carter_collisional_2018} showed that the masses of the largest remnant at both head-on and oblique impacts show excellent agreement with the updated universal law from \citet{leinhardt_collisions_2012}. Nonetheless, \citet{carter_collisional_2018} reported that the core mass fraction generally reaches a plateau at relatively high impact energies and deviates from the prediction laws proposed by \citet{marcus_collisional_2009}. However, very few hit-and-run simulations were performed in \citet{carter_collisional_2018}. \citet{gabriel_gravity-dominated_2020} also developed empirical relationships for the accretion and erosion of gravity-dominated bodies of various compositions during various impact conditions, but the largest target mass was around 1 M$_{\oplus}$. Studies by \citet{marcus_collisional_2009, leinhardt_collisions_2012, carter_collisional_2018, gabriel_gravity-dominated_2020} have mainly focused on low-velocity collisions and lack simulation data for extreme impact conditions where super-Mercuries could be formed. Additionally, all simulations except for those by \citet{marcus_collisional_2009} are designed to study smaller bodies than Earth, and focus more on the collision outcome of planetesimals.

As a replacement for the scaling laws proposed by \citet{marcus_collisional_2009}, \citet{reinhardt_forming_2022} proposed new scaling laws based on super-Earth head-on impact simulations with the new M-ANEOS equation of state \citep{stewart_equation_2019, stewart_equation_2020} at high enough impact velocities to form very dense planets. They showed that in the critical disruption regime, above a certain normalized impact energy, the mass of the largest remnant will rapidly decrease and the iron mass fraction will increase rapidly, both following a power law. These new scaling laws deviate significantly from those proposed by \citet{marcus_collisional_2009} and \citet{carter_collisional_2018} at high impact energies.

In addition to the scaling laws derived from planetesimal impacts and two-layer super-Earth impacts, \citet{denman_atmosphere_2020, denman_atmosphere_2022} conducted a comprehensive study on the loss of atmosphere during both head-on and oblique impacts between super-Earth-sized planets with thick atmospheres. They derived scaling laws to predict the total mass and remaining atmosphere mass of the largest remnant. Although the target planets used in their study had three layers and massive atmospheres, at high impact energies, most of the atmosphere is expelled, and the scaling law provides a useful reference for super-Earth impacts without atmospheres.

In this paper, we aim to bridge the gap between previous studies and provide a more systematic survey of giant impacts that could potentially form super-Mercuries. We present the results of over 1200 smoothed-particle hydrodynamics giant impact simulations, with target masses ranging from 0.06 M${_\oplus}$ to 20 M${_\oplus}$ and impact parameters from 0 (head-on) to 0.7 (45$^\circ$). We propose new scaling laws for both head-on and oblique impacts that can be used to constrain the collision conditions of super-Mercuries.

This paper is organized as follows. In Section \ref{sec:Methods}, we describe the methods used to create planetary bodies and their respective giant impact SPH simulations. We explain the tools and methods used to analyze the results of simulations. In Section \ref{sec:results}, we present the results of head-on and oblique impacts and the derived disruption criteria. In Section \ref{sec:discussion}, we discuss the applicability of our scaling laws to predict the outcome of giant impacts and the two important mechanisms involved in the mantle stripping process. In Section \ref{sec:conlcusion}, we summarize the results presented in the paper.

\section{Methods}
\label{sec:Methods}
\begin{figure}

	\includegraphics[width=\columnwidth]{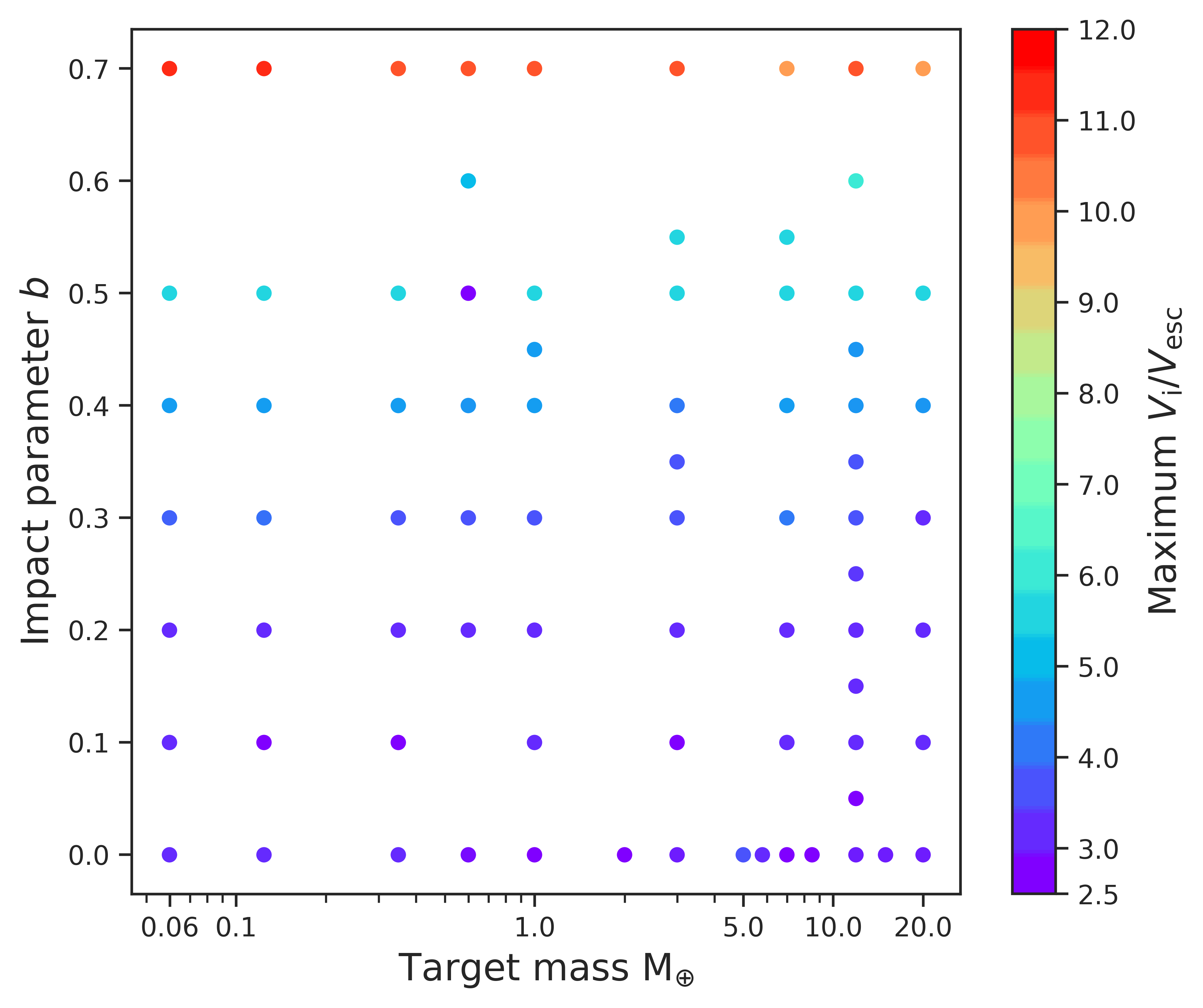}
    \caption{The parameter space of impact simulations in this work. The color of each point represents the maximum impact velocity normalized by the mutual escape velocity of a specific target mass and impact parameter.}
    \label{fig:impact_parameters}
\end{figure}

We ran around 1200 SPH simulations with various target masses, impact velocities, and impact angles using \texttt{SWIFT}(v 0.9.0) \citep{schaller_swift_2016,kegerreis_planetary_2019,schaller_swift_2023}. We focus only on equal-mass giant impacts as these are the most efficient collision type to form dense planets in terms of total impact energy both for head-on impacts and oblique impacts \citep{leinhardt_collisions_2012,denman_atmosphere_2020}. Thus, we set the impactor to target mass ratios ($\gamma=M_\mathrm{imp}/M_\mathrm{targ}$) to one for all simulations in this work. Figure \ref{fig:impact_parameters} shows the parameter space spanned in our simulations. We simulated targets spanning the range of rocky planet masses: $0.06$, $0.12$, $0.35$, $0.6$, $1.0$, $2.0$, $3.0$, $5.0$, $5.8$, $7.0$, $8.5$, $11.9$, $15.0$ and $20.0\,\mathrm{M_{\oplus}}$, mostly comprised of 2-5 $\times$ $10^5$ SPH particles. 
 
\subsection{Initial conditions}
\label{sec:ics} 
The planetary profiles and initial conditions were generated with \texttt{WoMa} \citep{kegerreis_planetary_2019,ruiz-bonilla_effect_2020} initializing each layer with an isentropic temperature profile. The initial planets were differentiated with 30\% iron core and 70\% forsterite mantle. We use the iron \citep{stewart_equation_2020} and the forsterite \citep{stewart_shock_2020} M-ANEOS equations of state and re-generated the EoS tables with higher maximum densities (100\,$\mathrm{g\,cm}^{-3}$ for forsterite and 200 \,$\mathrm{g\,cm}^{-3}$ for iron) using \citet{stewart_equation_2019} and \citet{stewart_equation_2020}. We chose entropies of the mantle to ensure the mantles are hot but entirely below the melt curve, which varied between 2800 J K$^{-1}$ kg$^{-1}$ and 3027 J K$^{-1}$ kg$^{-1}$. The core entropies were set to give iron close to the solidus with temperatures larger than mantle temperature at the core-mantle boundary, which were between 1700 J K$^{-1}$ kg$^{-1}$ and 1800 J K$^{-1}$ kg$^{-1}$.  

Before the impact simulations, each body was equilibrated in a cooling simulation for 20\,hr of simulation time in isolation to reach a stable hydrostatic state. During the first 10\,hr, the entropy of the core and mantle were fixed to the desired values at each time step (see \citealt{carter_collisional_2018}) in order to produce planets with isentropic layers. Over the next 10 hr, planets evolved without this additional damping, towards a hydrostatic profile. After the equilibration process, particles have a root-mean-squared velocity that is less than 1\% of the planet's escape velocity.  
 
We ran simulations for a range of impact parameters and target masses; additional head-on impact simulations were run for target masses of 2, 5, 5.8, 8.5, and 15 M$_\oplus$. 
For oblique impact simulations, the impact parameter ($b=\sin{\theta}$, where $\theta$ is the angle between the centers of the bodies and the velocity vector at the time of contact, (see \citealt{leinhardt_collisions_2012}) ranges from 0.1 to 0.7. The impact velocities range from one to 11 times the mutual escape velocity ($V_\mathrm{esc}$), which is defined by:
\begin{equation}
    V_\mathrm{esc}=\sqrt{\frac{2G(M_\mathrm{targ}+M_\mathrm{imp})}{R_\mathrm{targ}+R_\mathrm{imp}}},
	\label{eq:ves}
\end{equation}
where $G$ is the gravitational constant, and $M_\mathrm{targ}$, $R_\mathrm{targ}$ and $M_\mathrm{imp}$, $R_{\mathrm{imp}}$ are the mass and radius of the target and the impactor, respectively. For equal-mass oblique impacts with larger impact parameters, the required impact speed to generate a relatively dense planet (e.g. 50\% iron mass fraction) could be more than seven times the mutual escape velocity of the two-body system. 
In this work, higher impact velocities were chosen for larger impact parameters. Head-on impacts have impact velocities up to 3 $V_
\mathrm{esc}$, while impacts with $b=0.7$ can have impact velocities as high as 11 $V_\mathrm{esc}$. 

Bodies were initially separated by a distance such that contact occurs one hour after the start of the simulation to allow tidal deformation of the SPH planets \citep{kegerreis_atmospheric_2020}. For head-on impact simulations, the simulation time was approximately 15-20 hours. We observed that the properties of post-collision remnants became relatively stable after 6-10 hours of simulation time at the particle resolution we used. For oblique collisions, the simulations typically ran for 30 hours to distinguish between accretion impacts and erosive hit-and-run impacts. The simulations were run in cubic boxes with sides ranging from 1000 $\mathrm{R}_{\oplus}$ to 5000 $\mathrm{R}_{\oplus}$, depending on impact velocities. Any particles leaving the box were removed from the simulation.

SWIFT uses a parameter $h_\mathrm{max}$ (see SWIFT documentation\footnote{\url{https://swift.strw.leidenuniv.nl/docs/index.html}}) setting the maximal allowed smoothing length of SPH particles. 
Particles in the simulations can only have smoothing length less than $h_\mathrm{max}$, which will in turn set a density floor (minimum density of SPH particles). A maximum smoothing length or a minimum SPH density has a great influence on the post-collision disk \citep{hull_effect_2023} as relatively few particles are left in the disk region.  Most of the simulations in this study ($N = 2 \times 10^5$) used an $h_\mathrm{max}=0.2  \mathrm{R}_{\oplus}$ resulting in a density floor of $\sim0.1$ g cm$^{-3}$. To understand the influence of maximum smoothing length on post-collision remnants, we tested a subset of simulations with a larger $h_\mathrm{max}$ ranging from $0.5  \mathrm{R}_{\oplus}$ to $5  \mathrm{R}_{\oplus}$. The mass and iron fraction of the largest post-collision remnant agrees within 5\% for all the investigated  $h_\mathrm{max}$ values. The larger the $h_\mathrm{max}$ the longer a simulation will take to run. Since this study focuses on the large-scale properties of post-collision remnants (rather than the properties of the disk), our smaller value for $h_\mathrm{max}$ provides an acceptable balance between computing cost and accuracy.

Finally, we also investigated the dependence of numerical resolution on the mass and iron mass fractions of the largest post-collision remnants. The resolution tests focused on the lower and higher mass range as these are most susceptible to phase changes and are thus the most sensitive to numerical resolution. We tested target masses below 1 $\mathrm{M_{\oplus}}$ and above 15 $\mathrm{M_{\oplus}}$ with higher resolution ($10^6$ particles) and did not find significant resolution dependence for the properties of post-collision remnants which is consistent with \citet{meier_eosresolution_2021}. 

\subsection{Remnant properties}
\label{sec:analysis}
We used the same method \citep{dou_planetboundmass_2023} as in \citet{marcus_collisional_2009} and \citet{carter_collisional_2018} to determine particles bound to a post-collision remnant. Particles' potential and kinetic energies were first computed in relation to the seed particle that was closest to the potential minimum. Then, for the bound particles, the centre of mass position and velocity were computed and used as the seed for the next iteration. This process was then repeated for the remaining unbound particles until the calculation converged. We do not take account of the re-accretion of ejecta unbound to the largest post-collision body but bound to the star, as in the previous similar studies \citep{marcus_collisional_2009,leinhardt_collisions_2012,carter_collisional_2018,reinhardt_forming_2022}. The mass of the resulting iron core was found from the mass of iron particles in the largest remnant. The iron mass fraction of a remnant is the ratio between its iron mass and remnant mass.

We calculate the radii of post-collision planets corresponding to the radius of a super-Earth, using interior models with a iron mass fraction and mass determined as described in Section \ref{sec:analysis}. We assume that all bound particles (debris) will re-accumulate and that the planets will eventually become differentiated again after a long cooling process. \citet{benz_origin_2007} studied the long-term evolution of debris after potential Mercury-forming impacts using analytical calculations and found that around 40\% of particles would be re-accreted by Mercury after several Myr. Therefore, our simple assumption of no re-accumulation of unbound debris should provide a lower limit for the radii of super-Earths. The evolution of ejected debris will vary for different impact scenarios and planetary systems, and the inclusion of the effect of re-accumulation is beyond the scope of this work. For each post-collision remnant mass and iron mass fraction, we use the planet interior model code {\sc magrathea} \citep{huang_magrathea_2022} to calculate the final planet's radius. We use {\sc magrathea}'s default core and mantle equations of state as described in \citet{huang_magrathea_2022} (note these are different to the equations of state used for the SPH simulations). We use isentropic temperature gradients and set the surface temperature to 1000K; we find that the final radii vary little with the surface temperature.

\subsection{Catastrophic disruption criteria}
\label{sec:qrd}
Scaling laws are typically functions of the mass or iron mass fraction of the largest remnant and the normalized impact energies. When the impact energies from different target masses are normalized by the catastrophic disruption criteria, $Q^{*}_\mathrm{RD}$ (the impact energy needed to disperse half of the system's total mass), they can be effectively summarized using similar equations.
We follow the same variable definition and annotation as in \citet{leinhardt_collisions_2012}. The specific impact energy $Q_\mathrm{R}$ is defined by:
\begin{equation}
    Q_\mathrm{R}=0.5\mu\frac{V^2_{i}}{M_\mathrm{tot}},
	\label{eq:QR}
\end{equation}
where $M_\mathrm{tot}=M_\mathrm{targ}+M_\mathrm{imp}$ is the total mass of the system, $\mu=M_\mathrm{targ}M_\mathrm{imp}/M_\mathrm{tot}$ is the reduced mass, and $V_\mathrm{i}$ is the impact velocity.  Therefore, $M_\mathrm{lr}/M_\mathrm{tot} = 0.5$ (where $M_\mathrm{lr}$ is the mass of the largest post-collision remnant) for a collision with specific reduced impact energy equal to $Q^{*}_\mathrm{{RD}}$. For head-on impacts, to determine the critical specific impact energy for catastrophic disruption we use linear interpolation between the two data points that surround the specific impact energy where about half of the total colliding mass remains bound.

During head-on impacts, most materials from the target and impactor will be compressed and shocked, in the resulting collisions. At impact energies meeting the criteria for catastrophic disruption, half of the masses are lost due to the high pressure or shock propagation experienced during the impacts. However, in oblique impacts, especially equal-mass impacts, half of the system's mass can be lost during hit-and-run impacts, where the impactor deposits little energy into the target. A significant amount of mass can be lost due to the misalignment between the target and the impactor.
As a result, if traditional catastrophic disruption criteria are still used to describe oblique impact systems, the criteria may not accurately reflect the energy behavior of such systems. Therefore, for oblique impacts, we define a new disruption criterion $Q^{'*}_\mathrm{TD}$ which is the specific impact energy required to disperse half of the target's mass. For each group of target mass and impact parameter, the target catastrophic disruption criterion $Q^{'*}_\mathrm{TD}$ was estimated by fitting the equation:
\begin{equation}
\frac{M_\mathrm{lr}}{M_\mathrm{targ}} = c_1Q_\mathrm{R}^{c_2}+c_3,
\label{eq:QTD}
\end{equation}
with $c_1$, $c_2$ and c$_3$ as free parameters and using data points of $M_\mathrm{lr}$/$M_\mathrm{targ}$ between 0.8 and 0.2, and evaluating $Q_\mathrm{R}$ when $M_\mathrm{lr}$/$M_\mathrm{targ}$ is 0.5. Figure \ref{fig:m1d0_fitting} shows an example of fitting process when $M_\mathrm{targ}$ is 1 M$_{\oplus}$.

\subsection{Hit-and-run velocity}
\label{sec:method_hnr}

For equal-mass oblique impacts, especially at low impact parameters, we find it is unlikely for a "clean" hit-and-run impact (where target masses remain unchanged) to occur, i.e. the mass of the largest remnant is equal to the target mass ($M_\mathrm{lr}=M_\mathrm{targ}$ and $M_\mathrm{lr}/M_\mathrm{tot}=0.5$). As impact speeds increase from the mutual escape velocity, equal-mass oblique impacts undergo a sudden transition from accretion dominated impacts to erosive hit-and-run or graze-and-merge. The transition region is quite sensitive to the initial impact scenario and makes it challenging to determine the exact impact speed at which a ``clean'' hit-and-run occurs. We therefore check the collision outcome more closely for impacts located in this transition region by running the SPH simulations for a longer time, up to 200 hr. We then ran simple $N$-body simulations using the center of masses and velocities of bound remnants with a duration of up to ten years as an additional way to determine the outcome an impact. All the $N$-body simulations in this work were run using the REBOUND $N$-body code \citep{rein_rebound_2012} and integrated using the hybrid symplectic MERCURIUS integrator \citep{rein_hybrid_2019}. We take the median speed of the two data points between which the ratio of the largest remnant mass to total mass ($M_\mathrm{lr}/M_\mathrm{tot}$) drops from above 0.5 to below 0.5 as the hit-and-run velocity. We used a finer speed grid around the speed where the hit-and-run transition begins, and the difference between the speeds we use to calculate median speeds is no larger than 1 km\,s$^{-1}$.

\section{Results}
\label{sec:results}
In this study, we simulate equal-mass collisions between target planets ranging from 0.06 M$_{\oplus}$ to 20 M$_{\oplus}$, encompassing a range of impact angle from head-on to oblique impacts. Our focus is on high-energy erosive impacts, where more than half of the total material in the system is ejected for head-on impacts, and more than half of the target mass is ejected for oblique impacts. Overall, our simulations reveal that the mantle stripping process is primarily influenced by two mechanisms: vaporization-induced ejection and kinetic momentum transfer. Vaporization plays a crucial role in head-on and nearly head-on impacts, while kinetic transfer significantly affects oblique impacts at relatively large impact angles.

We introduce new scaling laws for head-on impacts that accurately captures the subtle changes in mass and iron mass fraction of the largest remnant, thereby demonstrating that collisions between multilayered planets should be modeled separately at different impact energy regimes to better represent the distinct shock experiences encountered by each layer. 
Regarding oblique impacts, we employ a new criterion for target catastrophic disruption to normalize impact energies and present new scaling laws that can predict the outcomes of oblique impacts at different angles. 
Additionally, we present target mass dependent equations that predict the onset of equal-mass hit-and-run impacts. Based on the new scaling laws for both head-on and oblique impacts, we derive the radii of post-collision remnants at different impact velocities for both types of impacts and illustrate the new mantle stripping curves alongside potential super-Mercuries in a mass-radius diagram.
\subsection{Head-on impact simulations}
\label{ref:head-on}

\begin{figure*}
\centering
\includegraphics[width=0.8\textwidth]{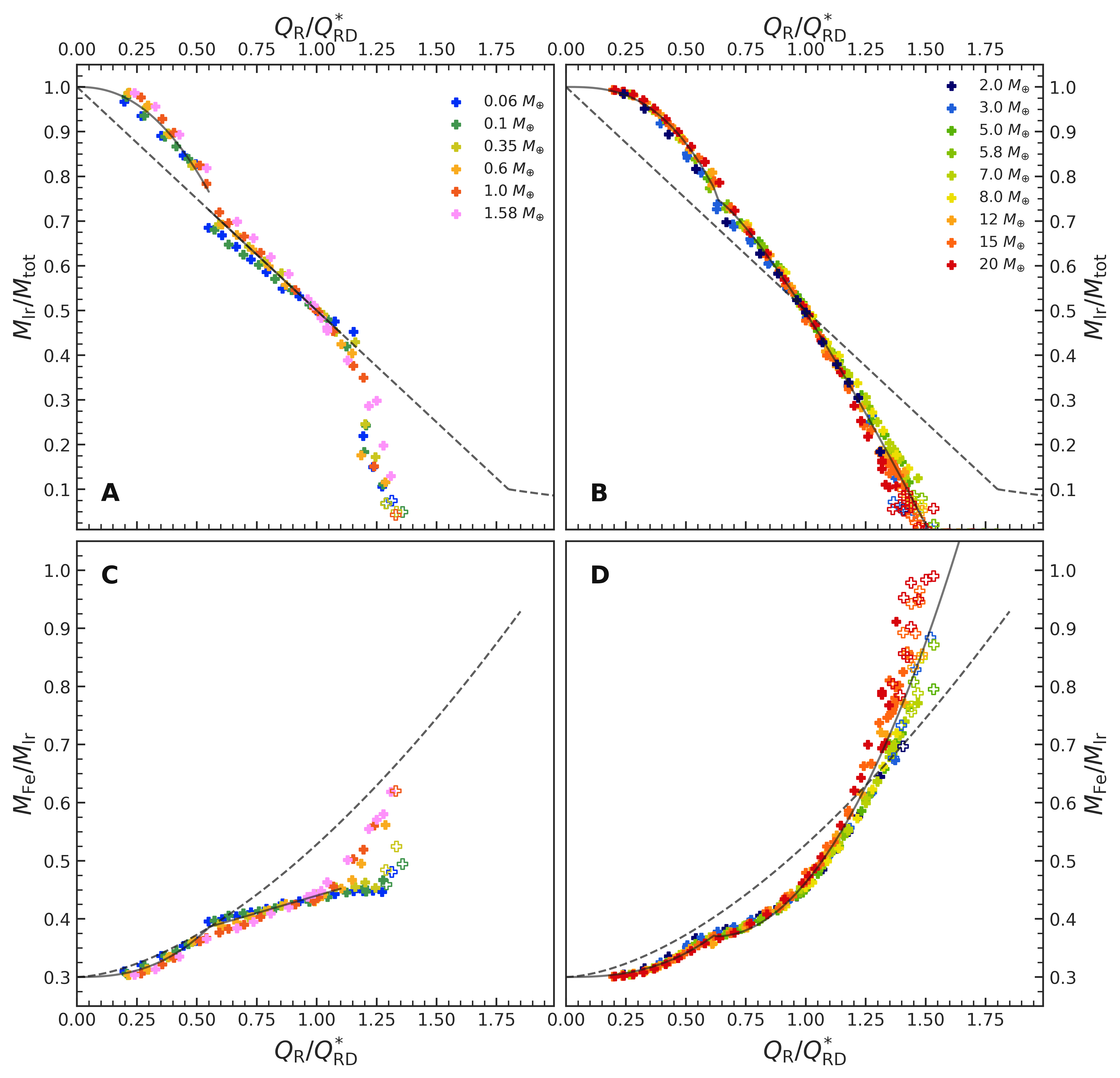}
\caption{Normalized mass (panel A and B) and iron mass fraction (panel C and D) of the largest post-collision remnant, plotted against the normalized impact energy $Q_\mathrm{R}$/$Q^*_\mathrm{RD}$ for all head-on collisions. Panels A and C display the results of impacts with target masses below 2 M$_{\oplus}$, while panels B and D show the impact results with target masses above 2 M$_{\oplus}$. The dashed lines in panels A and B represent the universal law from \citet{leinhardt_collisions_2012}, where there is a break in the slope at $Q_\mathrm{R}$/$Q^*_\mathrm{RD}$ equal to 1.8, above which is the super-catastrophic disruption power law. The dashed curves in panel C and D show the scaled down version of the fit from \citet{marcus_collisional_2009} updated by \citet{carter_collisional_2018}. The solid lines represent the scaling laws fitted from this work. Unfilled symbols show simulations with $M_\mathrm{lr}$/$M_\mathrm{tot}$ less than 0.1.}
\label{fig:head-on}
\end{figure*}

Figure \ref{fig:head-on} shows the mass and iron mass fraction of the largest remnant against the specific impact energy relative to the catastrophic disruption criteria for head-on impact simulations. Each symbol represents an impact simulation with different color for different target masses. We find that for head-on collisions, the impact results separate into two regimes at a target mass of $\sim 2 \mathrm{M}_{\oplus}$: the left hand panels (A \& C) in Figure \ref{fig:head-on} show low target mass impacts ($M_\mathrm{targ}<2 \mathrm{M_{\oplus}}$) and the right panels (B \& D) show high target mass impacts ($M_\mathrm{targ} \geq 2 \mathrm{M_\oplus}$). 

\subsubsection{Sharp change of mass and iron mass fraction}
\label{sharp_change}

$M_\mathrm{lr}$/$M_\mathrm{tot}$ shows several sharp changes in gradient: in panel A at around $Q_\mathrm{R}$/$Q^{*}_\mathrm{RD}$ $\simeq$ 0.55 and 1.15, and in panel B at around 0.625. In panels C and D, the increasing trend of iron mass fraction also changes at the normalized impact energies where the first sharp change in remnant mass occurs. 

In panel C, the iron mass fraction of low target mass impacts starts to increase more slowly above $Q_\mathrm{R}$/$Q^{*}_\mathrm{RD}$ $\simeq$ 0.55, 
and the maximum iron mass fractions of low target mass impacts are unlikely to exceed 50\%, which suggests a low mantle stripping efficiency for impacts in the low target mass regime. We note that \citet{carter_collisional_2018} found the iron mass fraction in their differentiated planetesimals impacts reached a plateau at a value of around 0.4, which is similar to iron mass fraction trend shown here in Figure \ref{fig:head-on}C. However, their break was around $Q_\mathrm{R}$/$Q^{'*}_\mathrm{RD}$=1.04 (\citealt{carter_collisional_2018} use $Q^{'*}_\mathrm{RD}$ to denote the catastrophic disruption threshold in their study) which is larger than the break 0.55 found in this work. The difference could be related to the different choice of initial core fraction of the target planets: the initial core fraction in \citet{carter_collisional_2018} is 22\% while our targets have a core fraction of 30\%.  According to \citet{leinhardt_collisions_2012, denman_atmosphere_2020, reinhardt_forming_2022}, head-on collisions should be the most efficient way to strip off mantle materials in terms of impact energy, but this might not be the case for low target mass impacts. Therefore, it is unlikely that Mercury and Mercury-like exoplanets that are small and dense with a core fraction larger than 50\% were formed by a single roughly equal mass head-on impact.

\begin{figure}
\centering
\includegraphics[width=0.9\columnwidth]{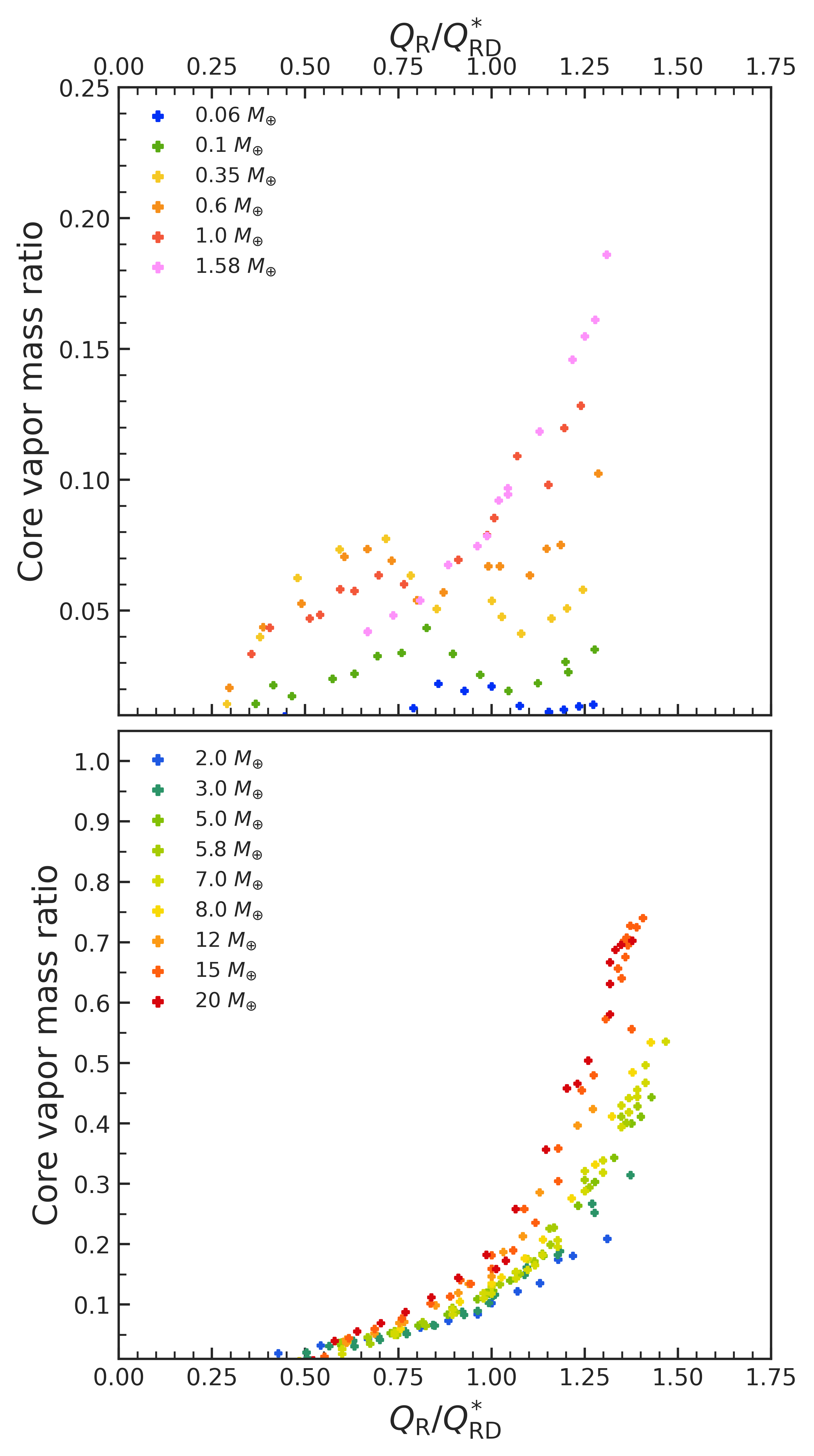}
\caption{Vapour fraction of the core material right after the end of each impact simulations. Top panel shows the vapour fraction of all the impacts with target mass below 2 M$_{\oplus}$, and bottom panel shows vapour fraction for impacts with target mass above 2 M$_{\oplus}$. Different colors represent different target mass in each panel. Note the scales of y axes of the two panels are different.}
\label{fig:vapour}
\end{figure}

At impact energies higher than the sharp change of $Q_\mathrm{R}$/$Q^{*}_\mathrm{RD}$ at 0.625 (Fig. \ref{fig:head-on}D), the iron mass fraction for high target mass impacts initially shows a slowly increasing trend similar to the low target mass impacts for a similar energy (Fig. \ref{fig:head-on}C). However, at energies greater than  $Q_\mathrm{R}$/$Q^{*}_\mathrm{RD}$ around 0.8-0.9 the iron mass fraction increases rapidly with increasing energy and can reach 80\% or greater. 

\subsubsection{Vaporization enhanced mantle stripping}
\label{vapor_stripping_headon}
The difference between the iron mass fraction of low and high target mass impacts at high impact energies is likely related to the vaporization of core materials. 
Figure \ref{fig:vapour} shows the vaporized fraction of all the core material at the final step of each impact simulation for the low and high target masses. The vapour fractions are calculated using the lever rule, and the entropies and pressures on the iron vapour dome are taken from \citet{stewart_equation_2020}. When normalized impact energy ($Q_\mathrm{R}$/$Q^{*}_\mathrm{RD}$) is small, the core vapour fraction of both low and high target mass impacts barely exceeds 10\%. As target mass increases, when $Q_\mathrm{R}$/$Q^{*}_\mathrm{RD}$ is larger than around 0.8-0.9, more than 10\% core materials in high target mass impacts can be vaporized as shown in the bottom panel of Figure \ref{fig:vapour}.

\begin{figure}
\centering
\includegraphics[width=0.9\columnwidth]{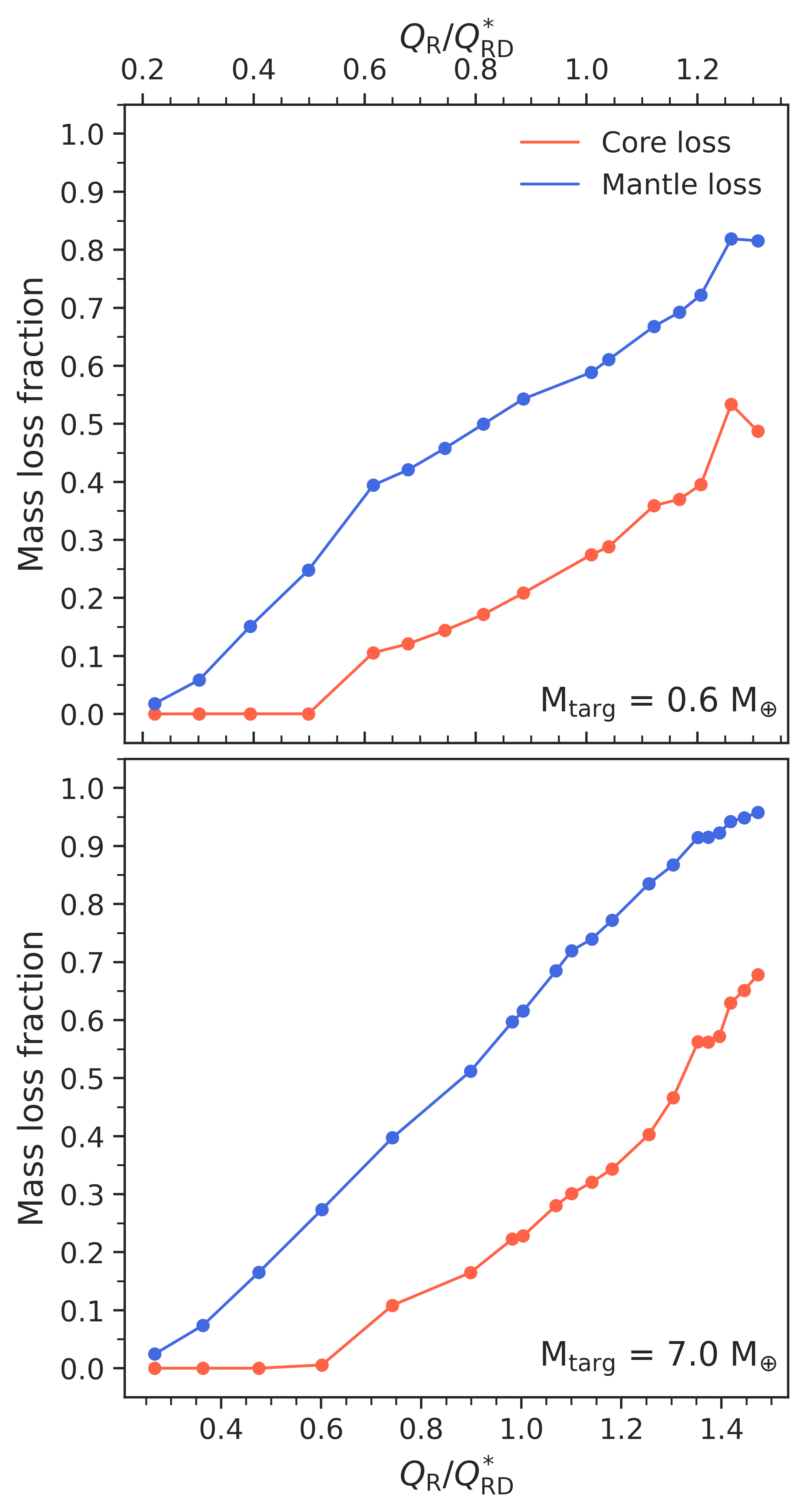}
\caption{The loss mass fraction of core (red dots and lines) and mantle (blue dots and lines) materials, plotted against the normalized impact energy for head-on collisions of target mass 0.6 M$_{\oplus}$ (top panel) and 7.0 M$_{\oplus}$ (bottom panel). }
\label{fig:core_mantle_loss}
\end{figure}

Therefore, we propose the mantle stripping process of differentiated planets impacts is dominated by two processes: kinetic momentum transfer and vaporization-induced ejection. At relatively low impact speeds, starting from $V_\mathrm{i}=V_\mathrm{esc}$, mantle materials begin to be expelled while only a small amount of core materials are kicked fast enough to overcome the gravitational binding energy of the system, since they are deeper and denser in the center of planets compared to mantle materials. With increasing impact energy, some core materials begin to be expelled, as shown by the red dots and lines in Figure \ref{fig:core_mantle_loss}. 
Core materials begin to be lost at around $Q_\mathrm{R}$/$Q^{*}_\mathrm{RD}$ = 0.5 for $M_\mathrm{targ}$ = 0.6 M$_{\oplus}$ (representing low target mass impacts) and $Q_\mathrm{R}$/$Q^{*}_\mathrm{RD}$ = 0.6 for M$_\mathrm{targ}$ = 7.0 M$_{\oplus}$ (representing high target mass impacts), which is consistent with the normalized energy where the first sharp change in remnant mass occurs (Fig. \ref{fig:head-on}). The iron mass fractions start to increase more slowly when core materials begin to be lost, since partial of the impact energy is deposited into core materials. If the core vaporization fractions remain very low, the increasing trend in iron mass fraction will appear as shown in Figure \ref{fig:head-on}C. In these simulations, we find that if the core vaporization fraction exceeds $\sim 10\%$, the vaporized core material will enhance the mantle stripping efficiency and make the iron mass fractions increase again, as shown in Figure \ref{fig:head-on}D.

\subsubsection{Fragmentation of the largest remnant}

\begin{figure*}
\includegraphics[width=\textwidth]{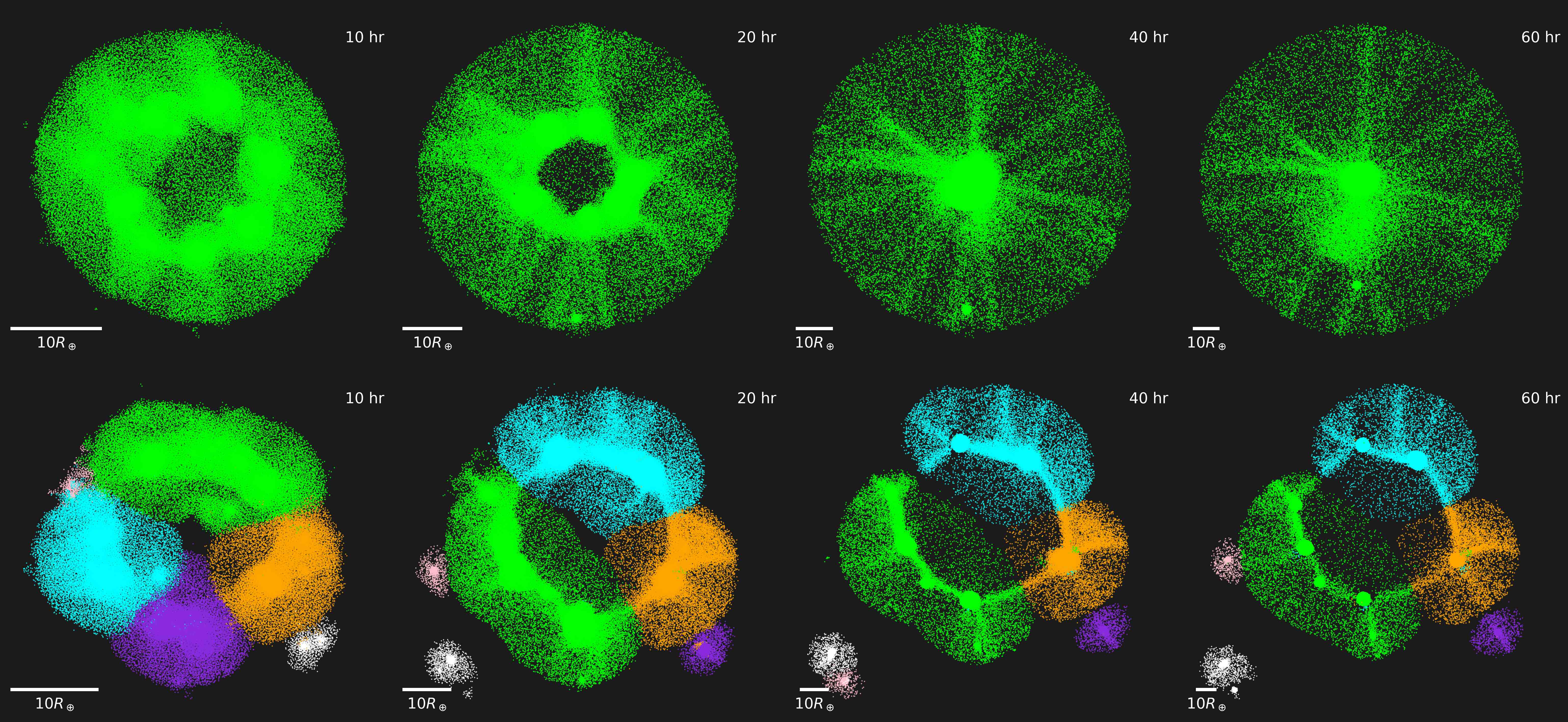}
\caption{Evolution of bound particles in the post-collision remnants after two equal mass 1.0 $\mathrm{M_{\oplus}}$ planets collided head-on at 30 km s$^{-1}$ (top panel) and 31 km s$^{-1}$ (bottom panel). Different colours represent different post-collision remnants. The green particles represent the largest remnant and the cyan particles are bound particles of the second largest remnant. In the bottom panel, from 10 hr to 20 hr, the largest remnant changes from the top to the lower left due to the merging of two remnants. }
    \label{fig:cs_combine}
\end{figure*}

Head-on impacts with low target masses will enter a fragmented regime when $Q_\mathrm{R}/Q^*_\mathrm{RD}$ is larger than approximately 1.15. 
In this regime, the post-collision largest remnant will split into several smaller pieces as shown in the bottom panel of Figure \ref{fig:cs_combine}. The fragmentation causes a rapid decrease in the largest remnant mass, as shown in Figure \ref{fig:head-on} panel A, around the second sharp change ($Q_\mathrm{R}/Q^*_\mathrm{RD}$ =1.15), which was also observed by \citet{reinhardt_forming_2022}. In Figure \ref{fig:cs_combine}, each color represents a bound remnant. In the top panel, for a head-on impact at velocity 30\,km\,s$^{-1}$ ($Q_\mathrm{R}$/$Q^*_\mathrm{RD}$ = 1.15), there is only one bound remnant. However, the bottom panel shows the bound remnant split into six smaller remnants at early times for the same mass impact with an impact speed of around 31 km s$^{-1}$ ($Q_\mathrm{R}$/$Q^{*}_\mathrm{RD}$ = 1.23).

We adopted several different approaches to verify this surprising fragmentation phenomenon. 
Firstly, we tracked the position of the remnants for ten years using simple $N$-body simulations (for details on code see section \ref{sec:method_hnr}), taking the center of mass and center of velocity of each remnant from the last snapshot (simulation time 60 hours) in the bottom panel and ignoring the effect of the central star. $N$-body simulations suggest that these remnants will move far away from each other and will not re-accrete to merge together again. Then, to check if the fragmentation is caused by the $h_\mathrm{max}$ or density floor set in the SWIFT simulations, we tested the same simulations with a larger $h_\mathrm{max}$ (up to 50 R$_{\oplus}$). We found that the fragmentation did not change which suggests it is not due to $h_\mathrm{max}$ being too small or the minimum resolved mass density being too large. As an additional comparison, using the same initial condition file and equation of state tables, we also performed equivalent impact simulations with the GADGET-2 SPH code \citep{cuk_making_2012} which does not have a density floor and found that similar fragmentation occurred  (although the mass of each remnant varied slightly -- within 10\% --from that found in the SWIFT simulations). Finally, we ran a friend-of-friend search \citep{creasey_tree-less_2018}  with a relatively large linking length (e.g., 5 R$_{\oplus}$) and found all particles were loosely bound together. We then applied our bound particle searching method to these particles to check if the fragmentation could be an artifact of the searching algorithm. Despite this, we still found the same splitting scenario with multiple remnants. Since all these tests show the fragmentation phenomenon, we infer that the  fragmentation is likely due to the equation of state used and might be related to the phase transition from liquid to vapour.

It is noteworthy that the fragmentation of the largest remnant does not only occur within the small target mass regime. Fragmentation occurs when $M_\mathrm{lr}$/$M_\mathrm{tot}$ is around 0.2 for target masses of 2.0 M$_{\oplus}$ and 3.0 M$_{\oplus}$. For target masses higher than 7.0 M$_{\oplus}$, fragmentation occurs when $M_\mathrm{lr}$/$M_\mathrm{tot}$ is less than 0.1, which is defined as the super-catastrophic disruption region \citep{leinhardt_collisions_2012}. Simulation data points of target masses 1.0 M$_{\oplus}$ (red dots) and 1.58 M$_{\oplus}$ (pink dots) in panels A and C of Figure \ref{fig:head-on} suggest that, with increasing target mass, the normalised energy required to trigger fragmentation also increases, and the mass and iron mass fraction of low target mass impacts will become increasingly similar to those of high target mass impacts. The fragmentation of the largest remnant in the small target mass regime might be crucial for the smaller planetesimal collision stage, where target masses are small but impact velocities are fairly large, which makes head-on or nearly head-on impacts more likely to happen. Previous N-body simulations have tended to use the universal law derived by \citet{stewart_velocity-dependent_2009} and \citet{leinhardt_collisions_2012} to predict the masses of remnants after collision. This law defines that fragmentation or super-catastrophic impacts only occur when $M_\mathrm{lr}$/$M_\mathrm{tot}$ is less than 0.1. However, based on our new simulations, fragmentation could occur very early when $M_\mathrm{lr}$/$M_\mathrm{tot}$ is as large as 0.4. We plan to conduct a more detailed study of the fragmentation in our following papers.

\subsubsection{Scaling laws}
We fit piecewise-defined scaling laws for the mass and iron mass fraction of the largest post-collision remnant separately for low and high target mass head-on impacts based on the locations of the first and second sharp changes.  For low target masses,

\begin{equation}
    \frac{M_\mathrm{lr}}{M_\mathrm{tot}}=
    \begin{cases}
    1 - 0.91 \Bigl(\frac{Q_\mathrm{R}}{Q^*_\mathrm{RD}}\Bigl)^{2.27} & \text{if }  \frac{Q_\mathrm{R}}{Q^*_\mathrm{RD}} \leq 0.55, \\
    \\
    0.75 - 0.50\Bigl(\frac{Q_\mathrm{R}}{Q^*_\mathrm{RD}}-0.50\Bigl) & \text{else}, \\
    \end{cases}
	\label{eq:head-on:mlr_lm}
\end{equation}

\begin{equation}
    \frac{M_\mathrm{Fe}}{M_\mathrm{lr}}=
    \begin{cases}
    0.3 + 0.39\Bigl(\frac{Q_\mathrm{R}}{Q^*_\mathrm{RD}}\Bigl)^{2.55} & \text{if} \frac{Q_\mathrm{R}}{Q^*_\mathrm{RD}} \leq 0.55, \\
    \\
    0.38 + 0.12 \Bigl(\frac{Q_\mathrm{R}}{Q^*_\mathrm{RD}}-0.5\Bigl) & \text{else}, \\
    \end{cases}
	\label{eq:head-on:zfe_l}
\end{equation}
and for high target masses,

\begin{equation}
    \frac{M_\mathrm{lr}}{M_\mathrm{tot}}=
    \begin{cases}
    1 - 0.86 \Bigl(\frac{Q_\mathrm{R}}{Q^*_\mathrm{RD}}\Bigl)^{2.77} & \text{if }  \frac{Q_\mathrm{R}}{Q^*_\mathrm{RD}} \leq 0.625, \\
    \\
    0.75 - 0.86 \Bigl(\frac{Q_\mathrm{R}}{Q^*_\mathrm{RD}}-0.625\Bigl)^{1.24} & \text{else}, \\
    
    \end{cases}
	\label{eq:head-on:mlr_hm}
\end{equation}

\begin{equation}
    \frac{M_\mathrm{Fe}}{M_\mathrm{lr}}=
    \begin{cases}
    0.3 + 0.25\Bigl(\frac{Q_\mathrm{R}}{Q^*_\mathrm{RD}}\Bigl)^{2.6} & \text{if} \frac{Q_\mathrm{R}}{Q^*_\mathrm{RD}} \leq 0.625, \\
    \\
    0.37 + 0.66 \Bigl(\frac{Q_\mathrm{R}}{Q^*_\mathrm{RD}}-0.625\Bigl)^{2}  & \text{else}. \\
    \end{cases}
	\label{eq:head-on:zfe_h}
\end{equation}
These scaling relations are shown using solid lines in Figure \ref{fig:head-on}.

For low target mass impacts, the first sharp change is around $Q_\mathrm{R}$/$Q^{*}_\mathrm{RD}$ equaling 0.55, and we only use simulation data where $Q_\mathrm{R}$/$Q^{*}_\mathrm{RD}$ is less than 1.15. Beyond this energy, impacts tend to become fragmented and $M_\mathrm{lr}$/$M_\mathrm{tot}$ shows a sudden drop from 0.4 to 0.1. For high target mass impacts, the sharp change occurs around $Q_\mathrm{R}$/$Q^{*}_\mathrm{RD}$ equaling 0.625, and we only use simulations where $M_\mathrm{lr}$/$M_\mathrm{tot}$ > 0.1, as impacts tend to be super-catastrophic below this mass and the largest remnant might not be resolved due to low number of particles. The boundary between low and high target mass scaling laws is not strict. Although fragmentation can still happen when the target mass is above 2.0 M${_\oplus}$, the fitted scaling laws can predict the mass of the largest remnant even after splitting when $M_\mathrm{targ}$ is above 2.0 M${_\oplus}$. Therefore, we recommend using high target mass scaling for $M_\mathrm{targ}$ equal to or above 2.0 M$_{\oplus}$ and low target mass scaling laws for target mass below 2.0 M$_{\oplus}$.

\subsection{Oblique impact simulations}

\begin{figure*}
    \centering
    \includegraphics[width=0.8\textwidth]{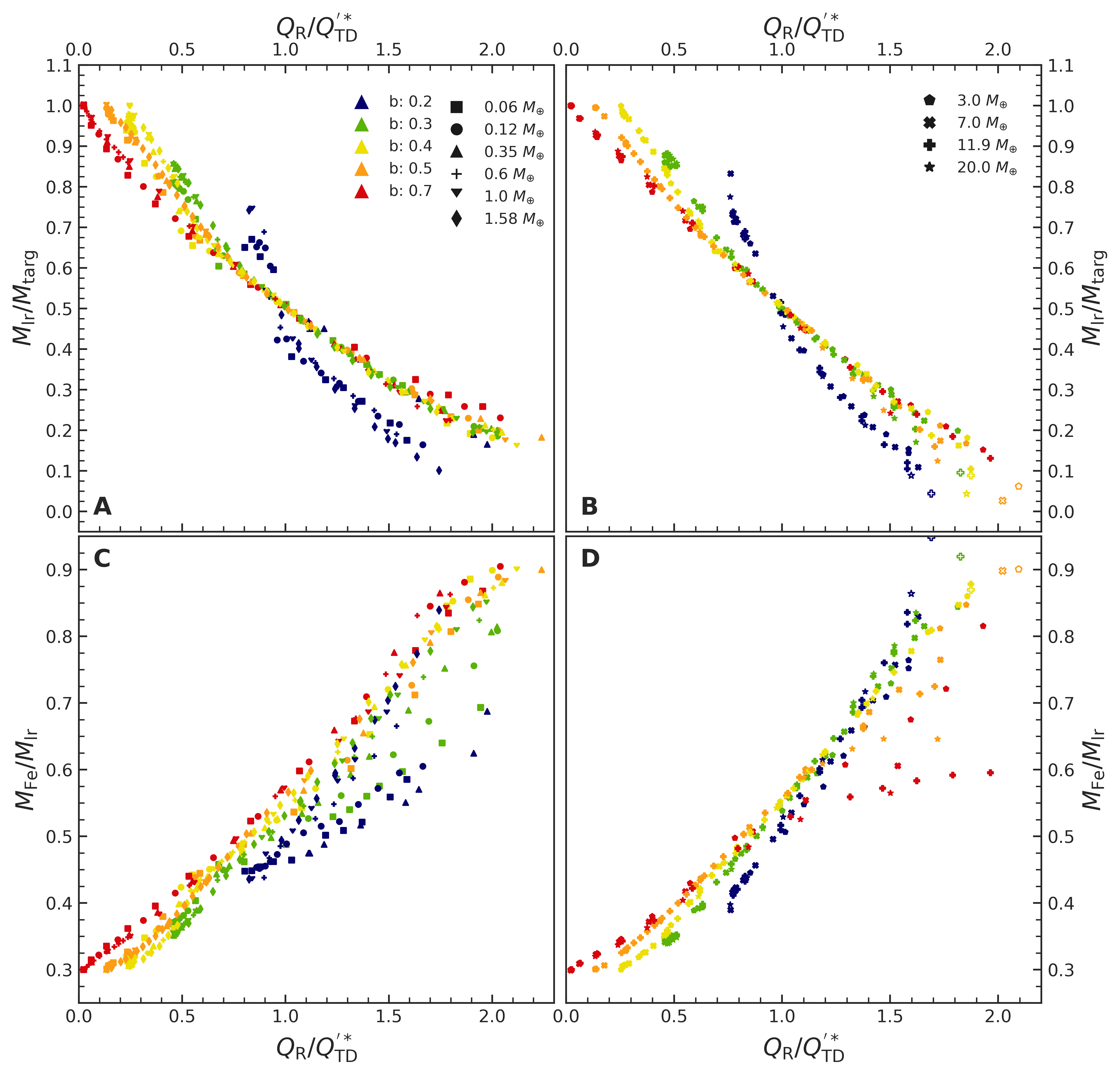}
    \caption{Normalized mass (panel A and B) and iron mass fraction (panel C and D) of the largest post-collision remnant, plotted against the normalized impact energy for all oblique collisions with low (left side panel A and C) and high (right side panel B and D) target masses. Unfilled symbols show the simulations with $M_\mathrm{lr}$/$M_\mathrm{targ}$ less than 0.1. }
    \label{fig:oblique}
\end{figure*}

Figure \ref{fig:oblique} presents the mass and iron mass fraction of the largest post-collision remnants versus normalized specific impact energies for oblique impacts. Note that for oblique impacts, we define target catastrophic disruption criteria, $Q^{'*}_\mathrm{TD}$, as the specific impact energy required to dispel half the mass of the target. Therefore, the mass of the largest remnant is now represented by $M_\mathrm{lr}$/$M_\mathrm{targ}$. We select and plot simulation data with remnant masses $M_\mathrm{lr}$/$M_\mathrm{targ}$ greater than 0.1 with filled symbols, below which the number of particles  in the post-collision remnants may not be sufficient for convergence. The left panels (A and C) show results with target mass less than 2.0 M$_{\oplus}$, and the right side panels (B and D) shows results for high target mass above 2.0 M$_{\oplus}$. Oblique impacts do not show a substantial difference in the mass of the largest remnant for different target mass regimes. However, we notice that high target mass impacts tend to have slightly lower remnant masses compared to low target mass impacts at higher impact energies with $Q_\mathrm{R}$/$Q^{'*}_\mathrm{TD}$ larger than 1.0.

In particular, at high impact energies with low impact parameters ($b=0.2$, blue symbols, and $b=0.3$, green symbols), the iron mass fractions of low target mass impacts (Figure \ref{fig:oblique}C) appear to be lower compared to those of higher target mass impacts, especially $M_\mathrm{targ}$ = 0.06 M$_{\oplus}$ and 0.12 M$_{\oplus}$ (squares and circles). Among all impact parameters tested, the mass and iron mass fraction of $b=0.2$ impacts follows a different pattern compared to others. Also, at $b=0.2$ and $b=0.3$, an impact tends to transition abruptly from accretion to erosive hit-and-run without going through a "clean" hit-and-run impact, and thus $M_\mathrm{lr}$/$M_\mathrm{targ}$ starts below 1.0 at these impact parameters. On the other hand, at $b=0.7$, impacts with target mass larger than 3.0 M$_{\oplus}$ at higher energies tend to have much lower iron mass fraction ($M_\mathrm{Fe}/M_\mathrm{lr}$ around 60\%) compared to low target mass high energy impacts ($M_\mathrm{Fe}/M_\mathrm{lr}$ around 90\%) as shown in Figure \ref{fig:oblique}D.

We fit scaling laws for erosive hit-and-run oblique impacts, as these are the impacts that can form dense planets. Impacts with velocities lower than the hit-and-run velocity can be approximated as perfect-merging events.  Therefore, only simulation data with $M_\mathrm{lr}$/$M_\mathrm{targ}$ less than 1.0 (beginning of erosive hit-and-run) and larger than 0.1 are selected to fit the scaling laws. The general equations for mass and iron mass fraction for oblique impacts are as follows:

\begin{equation}
    \frac{M_\mathrm{lr}}{M_\mathrm{targ}} =1.4\left(\alpha_{M,b} \right)^{\frac{Q_\mathrm{R}}{Q^{*}_\mathrm{TD}}}+(0.5-1.4\alpha_{M,b}),
	\label{eq:oblique-mlr}
\end{equation}

\begin{equation}
    \frac{M_\mathrm{Fe}}{M_\mathrm{lr}} =0.3+\alpha_{Fe,b}\left(\frac{Q_\mathrm{R}}{Q^{*}_\mathrm{TD}}\right)^{\beta_{Fe,b}}.
	\label{eq:oblique-zfe}
\end{equation}

\begin{figure}
    \centering
    \includegraphics[width=0.9\columnwidth]{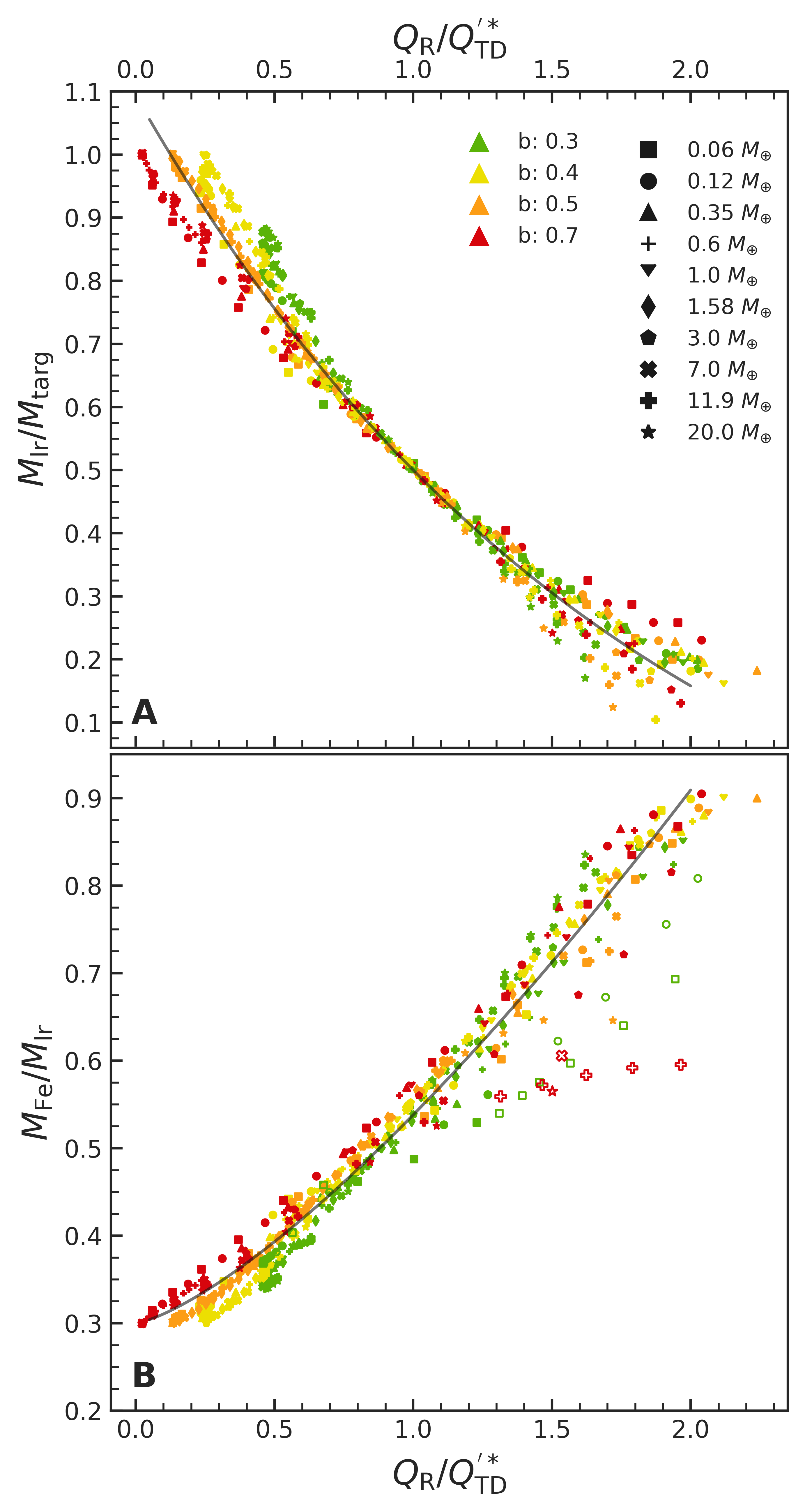}
    \caption{General fitted scaling laws of mass (top panel) and iron mass fraction (bottom panel) of the largest remnant for oblique impacts. Open symbols are excluded when fitting the scaling laws. Data with b=0.2 is not plotted.}
    \label{fig:ob_fit}
\end{figure}

Except for impacts at $b=0.2$, all impacts follow a similar trend both for mass and iron mass fraction. Thus, we first fit general scaling laws by combining all the simulation data from $b=0.3$ to $b=0.7$ as shown in Figure \ref{fig:ob_fit}, giving $\alpha_{M,b} = 0.576$, $\alpha_{Fe,b} = 0.238$, and $\beta_{Fe,b} = 1.356$ with a coefficient of determination $r^2$ of 0.973, and 0.976 for mass and iron mass fraction respectively. In order to balance the error especially for iron mass fraction, we exclude the iron mass fraction data points with $Q_\mathrm{R}$/$Q^{'*}_\mathrm{TD} >1.3$ for target mass 0.06 M$_{\oplus}$ and 0.12 M$_{\oplus}$ at $b=0.3$, and target mass larger than 3 M$_{\oplus}$ at $b=0.7$, which are shown as unfilled symbols in Figure \ref{fig:ob_fit}.

When $M_\mathrm{lr}$/$M_\mathrm{targ}$ is equal to 1.0, $Q_\mathrm{R}$/$Q^{'*}_\mathrm{TD}$ is not near zero, especially at low impact parameters. $M_\mathrm{lr}$/$M_\mathrm{targ}$ starts to change from greater than 1.0 to less than 1.0 when impacts transition from accretion to erosive hit-and-run. When $M_\mathrm{lr}$/$M_\mathrm{targ}$ is close to 1.0, the value of $Q_\mathrm{R}$/$Q^{'*}_\mathrm{TD}$ is determined by the hit-and-run velocity. For equal-mass oblique impacts, lower impact parameters tend to result in larger hit-and-run velocities (discussed in section \ref{sec:hnr}), and thus larger $Q_\mathrm{R}$/$Q^{'*}_\mathrm{TD}$ values. As a result, the data points for mass and iron mass fraction for each impact parameter slightly diverge from each other when $M_\mathrm{lr}$/$M_\mathrm{targ}$ is above $\sim$0.7. To more accurately model mass and iron mass fraction, we found that the parameters $\alpha_{M,b}$, $\alpha_{Fe,b}$, and $\beta_{Fe,b}$ in equations \ref{eq:oblique-mlr} and \ref{eq:oblique-zfe} can be described by the following equations at different impact parameters:
\begin{equation}
    \alpha_{M,b} = 1.602\log_{10}(b)-0.848b+1.478,
	\label{eq:amlr}
\end{equation}
\begin{equation}
    \alpha_{Fe,b} = 0.085b^2+0.218,
     \label{eq:afe}
\end{equation}
\begin{equation}
    \beta_{Fe,b} = 2.5\times(0.1)^b+b,
     \label{eq:bfe}
\end{equation}
where $b$ is the impact parameter between 0.3 and 0.7. Since impacts at $b=0.2$ are in the transition regime, and have relatively different trends of mass and iron mass fraction, we fit scaling laws at $b=0.2$ individually. At $b=0.2$, the mass follows
\begin{equation}
    \frac{M_\mathrm{lr}}{M_\mathrm{targ}} =2.8\times\left(0.23 \right)^{\frac{Q_\mathrm{R}}{Q^{*}_\mathrm{TD}}}+ (0.5-2.8\times0.23),
	\label{eq:oblique_fit_b0d2}
\end{equation}
and the iron mass fraction follows the same form as equation \ref{eq:oblique-zfe} with $\alpha_{Fe,0.2}=0.199$, and $\beta_{Fe,0.2}=1.904$. Iron mass fraction data points for target masses below 0.6 M$_{\oplus}$ are excluded from fitting at $b=0.2$.

\subsection{Catastrophic disruption criteria}
\begin{figure}
\centering
\includegraphics[width=0.9\columnwidth]{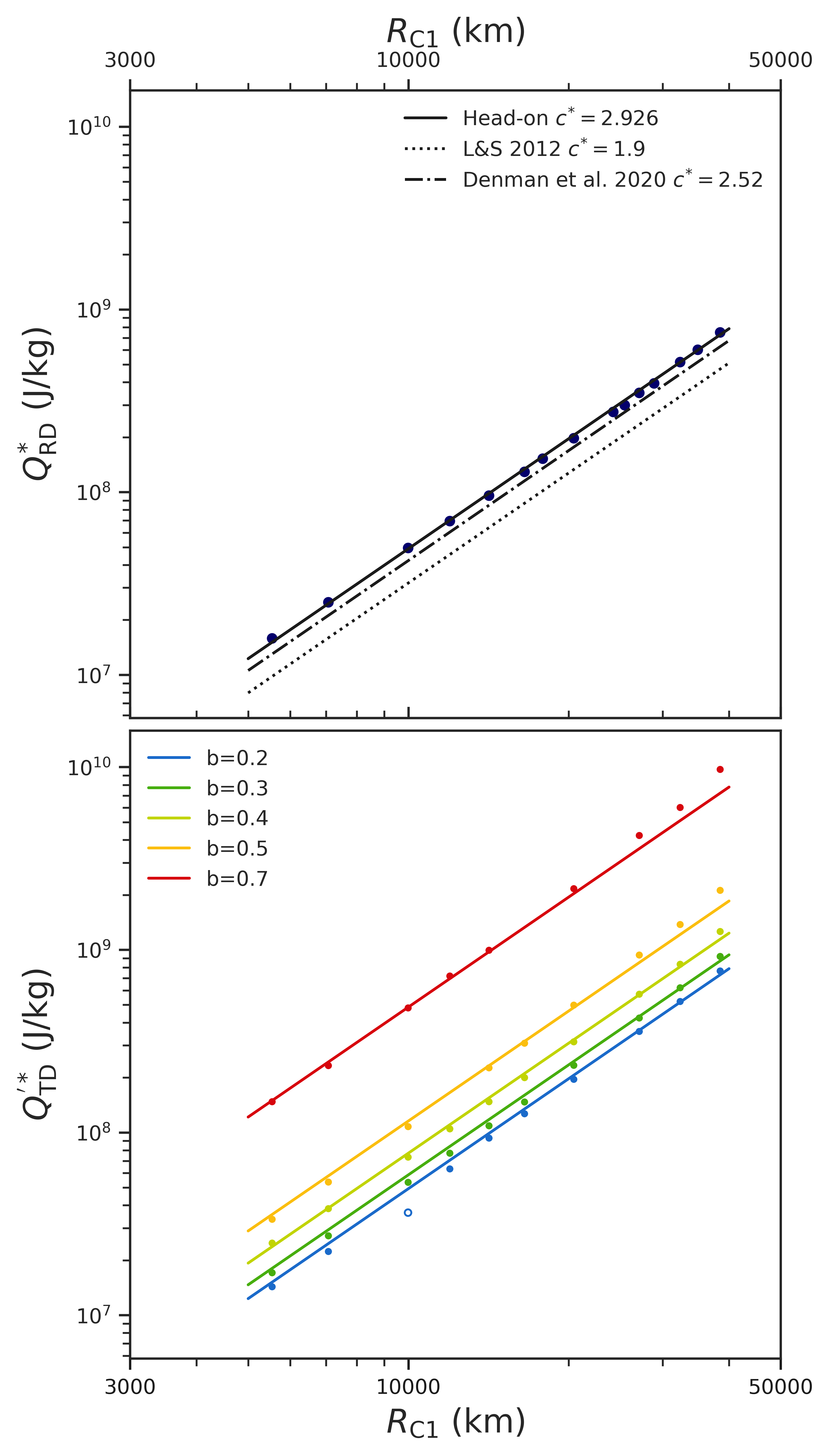}
\caption{Relationship between R$_{\mathrm{C1}}$ and catastrophic disruption criteria $Q^{*}_\mathrm{RD}$ for head-on impacts (top panel) and target catastrophic disruption criteria $Q^{'*}_\mathrm{TD}$ for oblique impacts (bottom panel).The Unfilled symbol at b=0.2 and M$_\mathrm{targ}$ =0.35 M$_{\oplus}$ was excluded from fitting.}
\label{fig:qrd_criteria}
\end{figure}

\citet{leinhardt_collisions_2012} reported that, for equal-mass head-on impacts, the catastrophic disruption criteria or the principal disruption curve follows: 
\begin{equation}
    Q^{*}_{\mathrm{RD},\gamma=1} = c^{*}\frac{4}{5}\pi\rho_{1}GR^2_\mathrm{C1},
    \label{eq:c-star} 
\end{equation}
where $c^*$ is a measure of the catastrophic disruption threshold in units of the gravitational binding energy, $\rho_1$ = 1000 kg m$^{-3}$, and $R_\mathrm{C1}$ = $(\frac{3 M_\mathrm{{tot}}}{4\pi\rho_1})^{\frac{1}{3}}$ is the radius a spherical body would have if it had the total system mass and a density of $\rho_1$. \citet{leinhardt_collisions_2012} found $c^*$ to be 1.9$\pm$0.3 (for planetary mass bodies), while \citet{denman_atmosphere_2020} fitted $c^*$ for planets with atmospheres to be around 2.52. Based on all of our head-on impact data, we found a $c^*$ value of 2.926. The top panel of Figure \ref{fig:qrd_criteria} shows the relationship between $Q^{*}_\mathrm{RD,\gamma=1}$ and R$_\mathrm{C1}$, along with the fitting lines from our work and previous work. 

In \citet{denman_atmosphere_2020}, the target planets have thick and massive atmosphere layers, which can account for up to 30\% of the planet's total mass. For planets with the same mass, including an atmosphere increases the planet's radius and decreases the total gravitational binding energy. This means that less impact energy is required for catastrophic disruption. This explains why their value of $c^*$ is slightly lower than ours. 

The difference between our results and those of \citet{leinhardt_collisions_2012} may also be related to the different internal structures of the target planets used in the simulations. \citet{leinhardt_collisions_2012} fitted their $c^*$ by combining previous hydrodynamic simulation results, including collisions between differentiated iron-rock bodies, differentiated ice-rock bodies, and pure rock bodies. The lower $c^*$ value, compared to the fitted $c^*$ from iron-rock bodies of this study, is a result of combining a range of different density planets, as ice-rock and pure rock bodies of the same masses have lower gravitational binding energies. In addition, the choice of equations of state, and number of fitting points used may also affect the final result. Our simulation data can be directly used for fitting since they are all equal-mass impacts. However, in \citet{leinhardt_collisions_2012}, a correction related to the mass ratio, $\gamma$, needs to be applied first to the raw data to convert the results to the equal-mass equivalent, which necessarily introduces some errors. The new M-ANEOS equation of state tables used in this work are better at modeling the phase transition from liquid to vapor state, and therefore will have a more accurate representation around the catastrophic disruption criteria. 

For oblique impacts with the same mass, we found that the newly defined target catastrophic disruption criteria $Q^{'*}_\mathrm{TD}$, which represents the specific impact energies required to expel half of the target's mass, follows a similar trend as equation \ref{eq:c-star} for head-on impacts but with larger constant parameters. $Q^{'*}_\mathrm{TD}$ at different impact parameter can be modeled as:
\begin{equation}
    Q^{'*}_\mathrm{TD} = \left(0.175 \left(\frac{1}{1-b} \right)^{3.3}+0.635\right)c^* \frac{4}{5}\pi\rho_{1}GR^2_\mathrm{C1},
    \label{eq:Q_td}
\end{equation}
where $c^*$ is the same as fitted for head-on impacts. The bottom panel in Figure \ref{fig:qrd_criteria} shows target catastrophic disruption criteria for oblique impacts and the corresponding fitting lines. 

\subsection{Hit-and-Run velocities}
\label{sec:hnr}
\begin{figure}
\centering
\includegraphics[width=0.9\columnwidth]{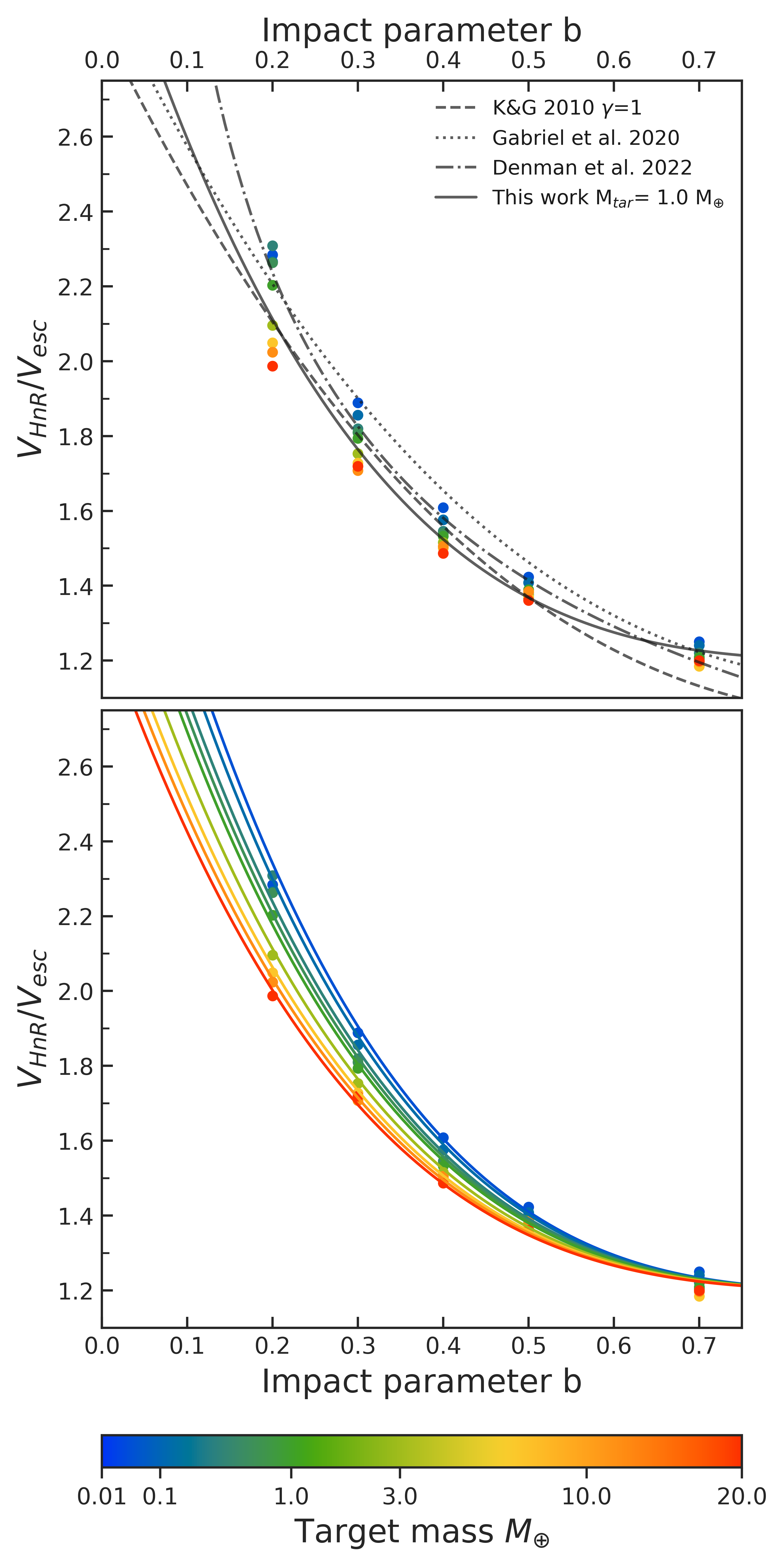}
\caption{The relationship between normalized hit-and-run velocities and impact parameter. $V_\mathrm{esc}$ is the mutual escape velocity. Colours represent the mass of the target in Earth masses. Top panel:The dashed line is the hit-and-run velocity criteria of \citet{kokubo_formation_2010} at $\gamma$=1, the dash dot line is the criteria from \citet{denman_atmosphere_2022}, and the dotted line is the fitting results from \citet{gabriel_gravity-dominated_2020}. Our fitted criteria at M$_\mathrm{targ}$ =1 M$_{\oplus}$ is the solid black line.  Bottom panel: Solid coloured lines show the target mass dependent hit-and-run criteria described by equation \ref{eq:v_hnr}.}
\label{fig:v hnr}
\end{figure}

Figure \ref{fig:v hnr} shows the normalized velocities where hit-and-run impacts begin to happen for different impact parameters for our equal mass oblique impacts. Hit-and-run velocities become more target mass dependent with decreasing impact parameter. Previously published hit-and-run criteria, as shown in the top panel of Figure \ref{fig:v hnr}, fit the general trend of hit-and-run velocities relatively well, but can not account for the variation with target mass. 

The dependency on target mass at low impact parameter is related to the core radius ratio, $R_\mathrm{core}$/$R_\mathrm{planet}$. The core radius ratio of each target planet decreases with increasing mass, for example, $R_\mathrm{core}$/$R_\mathrm{planet}$ is 0.52 for a target planet with a mass of 0.06 M$_{\oplus}$, while it is 0.48 for a target with a mass of 20 M$_{\oplus}$. In an impact where the target and impactor are the same size, the impact parameter at which the core material from both planets no longer intersect along the impact direction is the core radius ratio ($b=$ $R_\mathrm{core}$/$R_\mathrm{planet}$, if the deformation of the planets when they approach each other is not accounted for, Figure \ref{fig:impact_geo}). Below this limit, the projected length of the impactor's core overlapping the target's core is 
\begin{equation}
    l_\mathrm{core} =  2R_\mathrm{planet} \left(\frac{R_\mathrm{core}}{R_\mathrm{planet}}-b \right).
    \label{eq:core_ratio}
\end{equation}
Therefore, at the same impact parameter, lower mass targets have slightly larger core overlap and thus require higher impact velocities to transition into hit-and-run. 

We fit the hit-and-run velocity as a function of target mass as shown in the bottom panel of Figure \ref{fig:v hnr}. We do not include the data for $b=0.1$, as the impact angle is too small and the tidal deformation of the planets makes impacts behave very similarly to head-on impacts. Our target mass dependent hit-and-run criteria is 
\begin{equation}
    \frac{V_\mathrm{HnR}}{V_\mathrm{esc}} = \Gamma(1-b)^{3.6}+1.2,  
    \label{eq:v_hnr}
\end{equation}
where $\Gamma = -0.3\log_{10} M_{\mathrm{targ},\oplus} + 2.18$, and $M_{\mathrm{targ},\oplus}$ is the mass of target planet in Earth masses. Since all our simulations have impactor to target mass ratio $\gamma=1$, the hit-and-run criteria does not account for the effect of varying $\gamma$.

\section{DISCUSSION AND CONCLUSIONS}
\label{sec:discussion}

\subsection{Prediction of impact conditions}

\begin{figure*}
    \centering
    \includegraphics[width=0.8\textwidth]{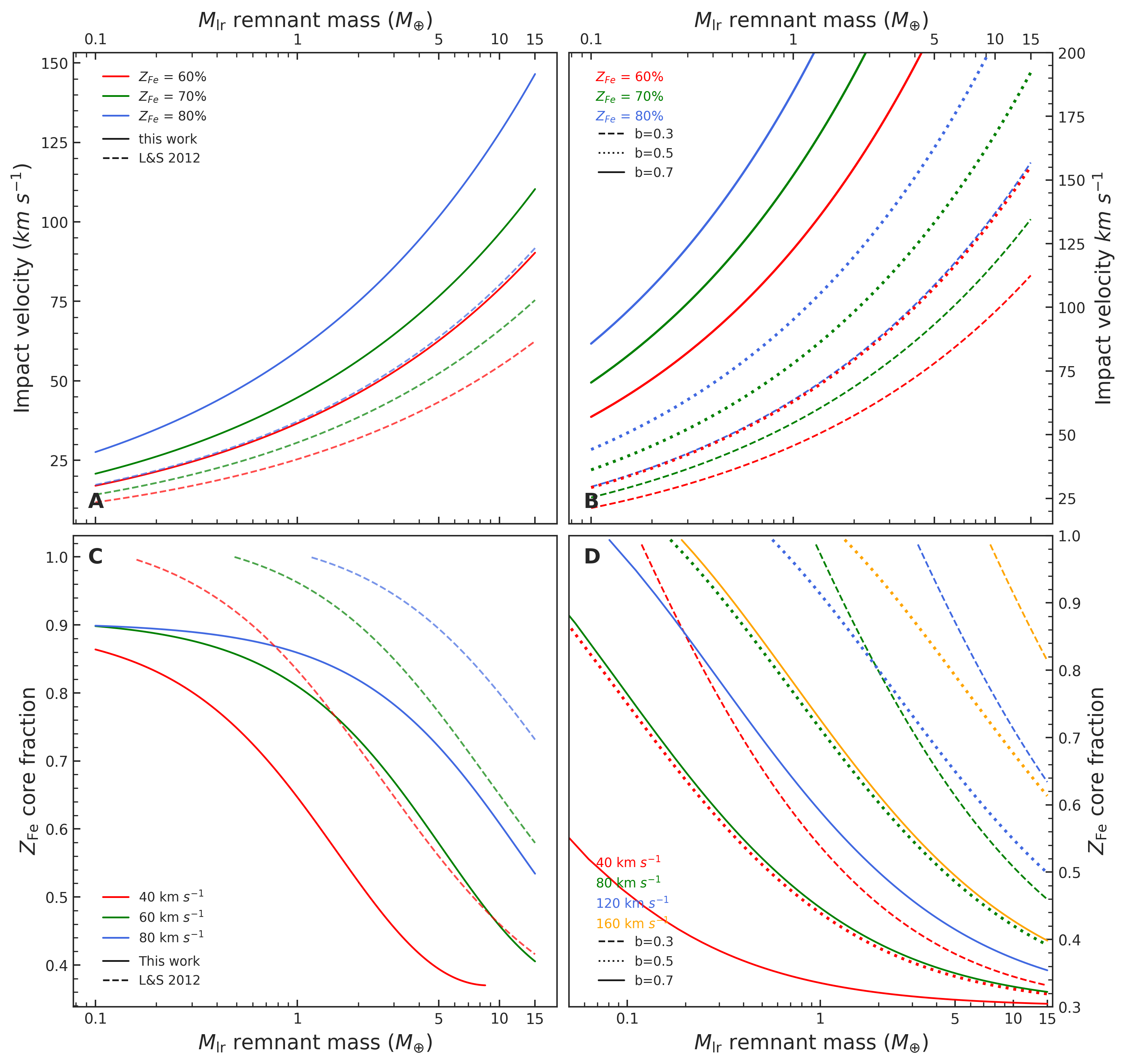}
    \caption{Prediction of impact conditions for head-on impacts (left panels A and C) and oblique impacts (right panels B and D). Top panels A and B: the impact velocity versus the mass of the largest remnant for iron mass fraction of 60\%, 70\%, and 80\%. Bottom panels C and D: the iron mass fraction versus the mass of the largest remnant for different impact velocities. Dashed lines show the prediction derived by combination of \citet{leinhardt_collisions_2012} and \citet{marcus_minimum_2010}.}
    \label{fig:v_z_m}
\end{figure*}

Prediction of impact conditions required to achieve a particular mass and core fraction can be made by combining the mass and iron mass fraction scaling laws derived in the previous sections. As we are motivated to model the formation of dense planets, for the head-on scaling laws, we only use the high target mass scaling law equations \ref{eq:head-on:mlr_hm} and \ref{eq:head-on:zfe_h} when $Q_\mathrm{R}$/$Q^{*}_\mathrm{RD}$ is larger than 0.625 because low target mass head-on impacts can at most form a remnant with 50\% iron mass fraction. Applying our new $c^* = 2.926$ and combining equations \ref{eq:head-on:mlr_hm}, \ref{eq:head-on:zfe_h} and \ref{eq:c-star}, Figure \ref{fig:v_z_m} A and C shows the prediction of head-on impact velocities and iron mass fractions derived from both our scaling laws and the combination of those from \citet{leinhardt_collisions_2012} and \citet{marcus_minimum_2010}. For the same remnant mass and iron mass fraction, we require higher impact velocities compared to \citet{leinhardt_collisions_2012} as our $c^*$ are larger. At different impact velocities, our iron mass fraction predictions converge at around 0.9. 
From the prediction of this work, the impact velocities necessary to produce a super-Mercury with mass above 5 M$_{\oplus}$ and iron mass fraction larger than 60\% are greater than 60 km s$^{-1}$, which suggests an impact occurring close to the star.

\begin{figure}	
\includegraphics[width=\columnwidth]{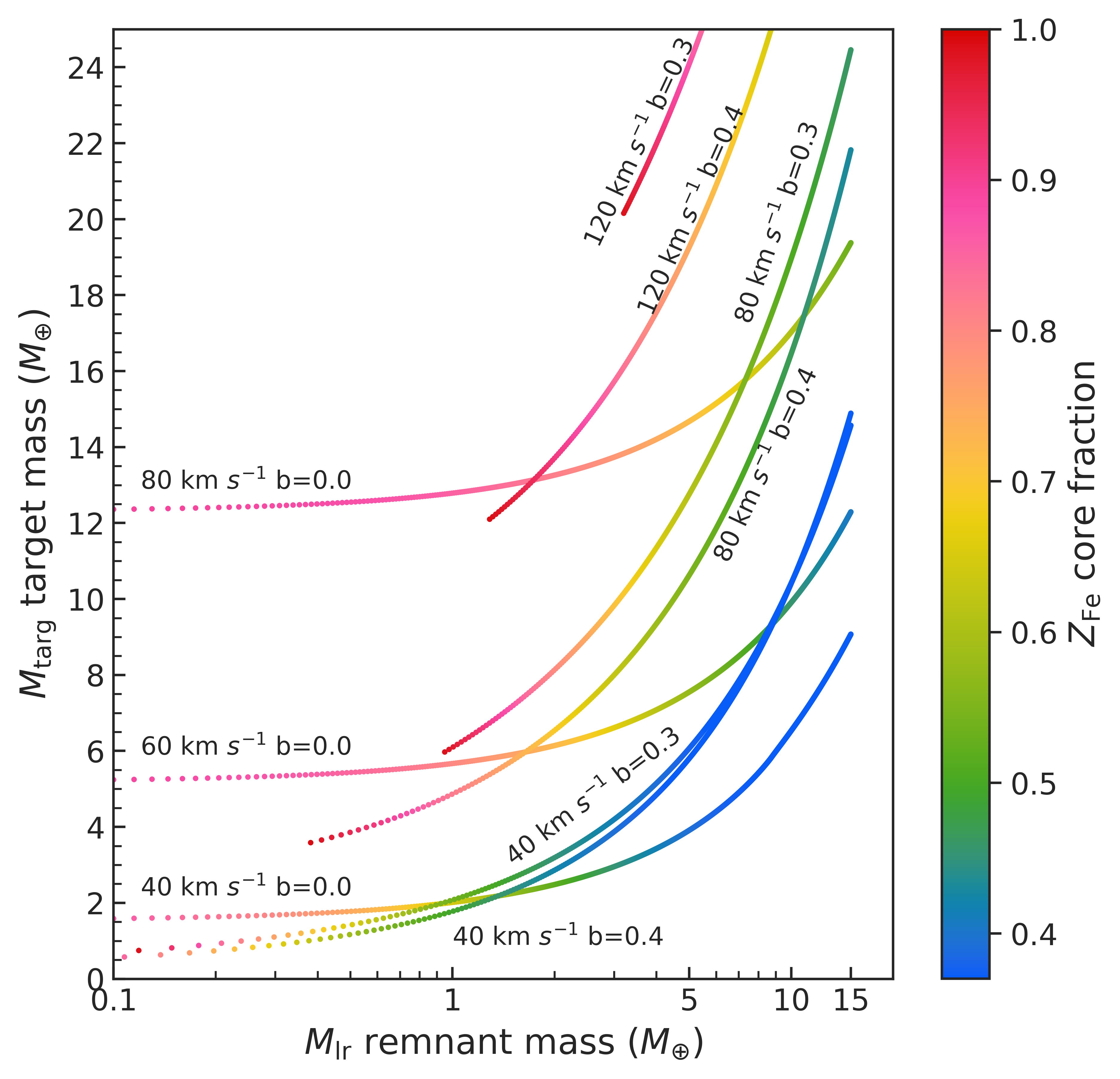}
    \caption{The relationship between the mass of the largest remnant and the inferred target mass at different impact velocities. Here, we present predictions for impact parameters of b =0.0, b=0.3, and b=0.4.}
    \label{fig:mtar_mlr}
\end{figure}

\begin{figure}	
\includegraphics[width=\columnwidth]{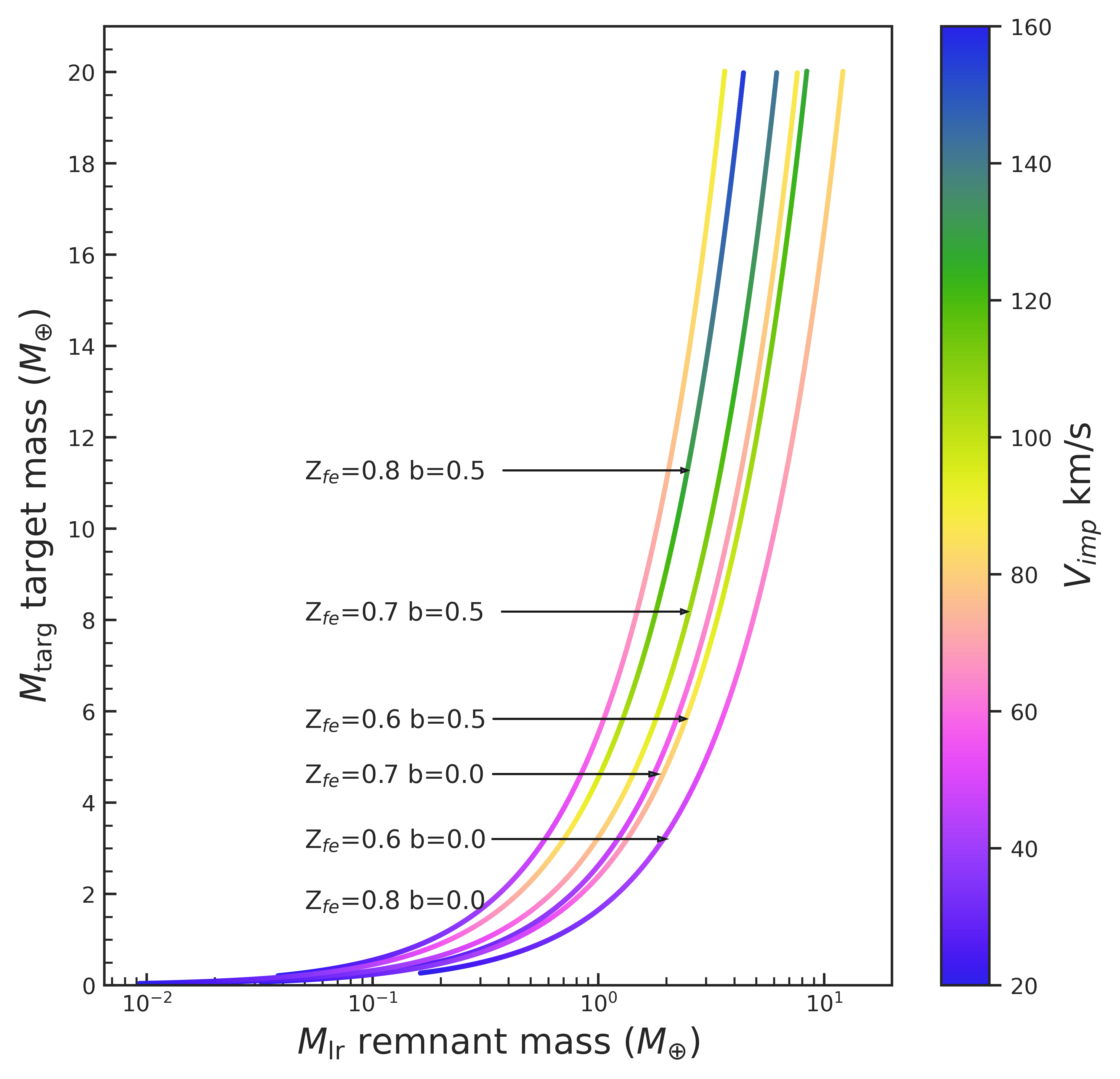}
    \caption{The relationship between the mass of the largest remnant and the inferred target mass at different iron mass fraction. To avoid overlap of lines, we present only predictions for impact parameters of b=0.0 and b=0.5.}
    \label{fig:mtar_mlr_zfe}
\end{figure}

For oblique impacts, combining equations \ref{eq:oblique-mlr}, \ref{eq:oblique-zfe}, and \ref{eq:Q_td} in a similar way, we derive the prediction of impact conditions for $b=0.3$, 0.5 and 0.7 shown in panel B and D in Figure \ref{fig:v_z_m}. For the same remnant mass and iron mass fraction, larger impact parameter requires higher impact velocities. At $b=0.7$, a remnant with mass above 5 M$_{\oplus}$ and 60\% iron mass fraction would need an impact velocity greater than 200 km s$^{-1}$. As shown in the Figure \ref{fig:v_z_m}D, for a remnant mass of 5 M$_{\oplus}$ and a velocity of 80 km s$^{-1}$, an impact at $b=0.7$ could only produce an iron mass fraction less than 40\%; while at $b=0.3$, the remnant could have an iron mass fraction of around 65\%. With increasing impact parameter, the mantle stripping efficiency drops significantly. 

The scaling laws can also be used to predict the pre-impact target masses required. For an exoplanet with a given mass and core fraction, Figure \ref{fig:mtar_mlr} provides a good reference for potential target masses and impact velocities that could form the planet via a giant impact. Figure \ref{fig:mtar_mlr_zfe} illustrates an alternative prediction of the mass of target bodies based on the iron mass fraction at different impact velocities. Comparing it to oblique impacts with $b$ equal to 0.5, head-on impacts require a higher target mass but lower impact velocities for a remnant with a specific mass and iron mass fraction.

\begin{figure}
\centering
% change legned of marcus
\includegraphics[width=1.0\columnwidth]{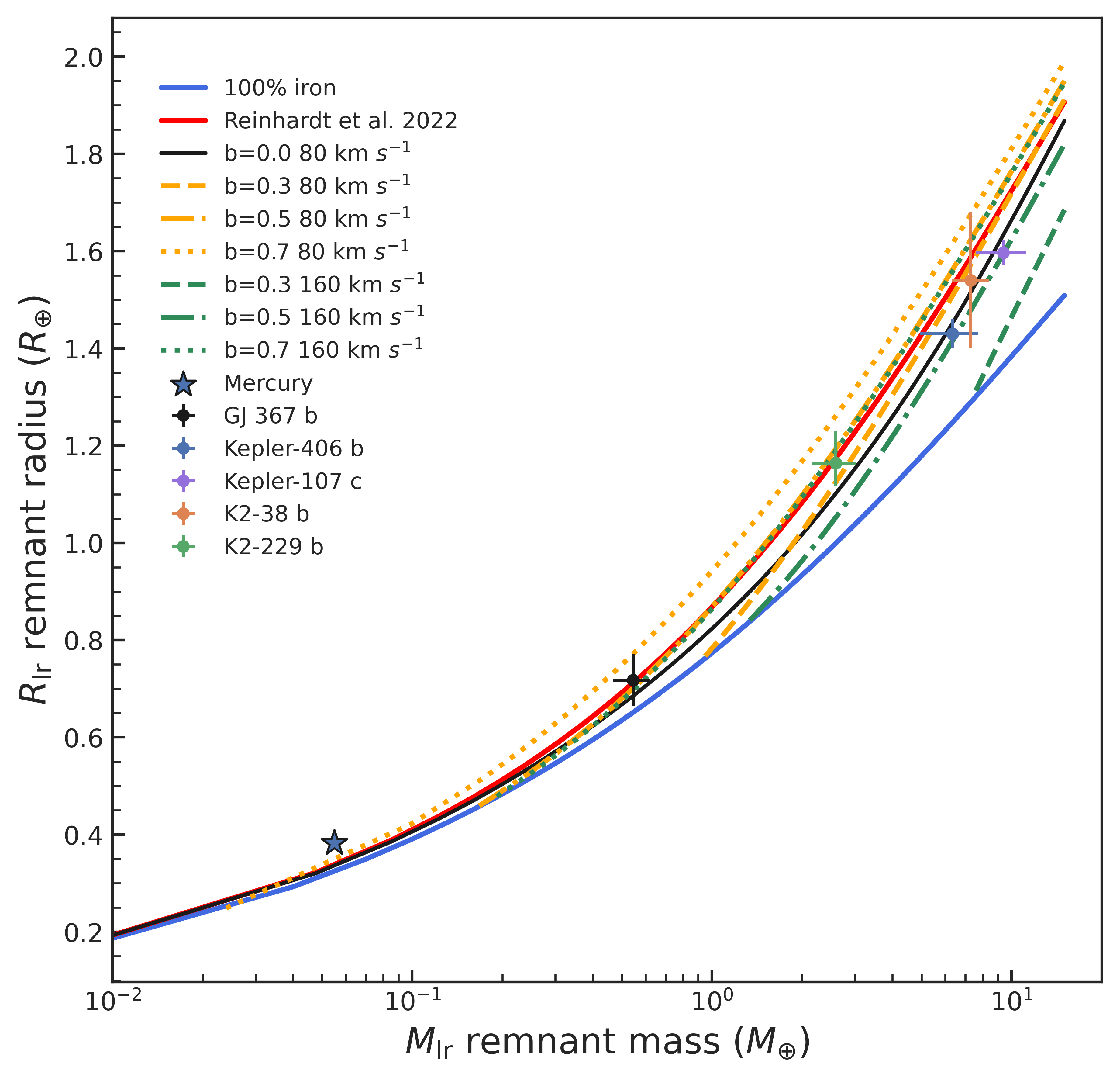} 
\caption{Mass-radius diagram for Mercury and super-Mercuries. The red solid line shows the \citet{reinhardt_forming_2022} corrected maximum stripping line generated using the method described in the section \ref{sec:analysis} for head-on impacts at 80 km s$^{-1}$ and the black solid line shows our new maximum stripping line. The dashed lines show the maximum stripping lines of oblique impacts. Blue solid line represent radii of plants made by 100\% core generated using the same method described in the section \ref{sec:analysis}.The data points show several possible super-Mercuries, GJ367\,b, Kepler-406\,b, Kepler-107\,c, K2-38\,b, and K2-229\,b.}
\label{fig:max}
\end{figure}

Using the method discussed in section \ref{sec:analysis}, we derived the radius\protect\footnotemark of post-collision remnants at different impact velocities for both head-on and oblique impacts, as shown in Figure \ref{fig:max}. These radii do not correspond to the radii of the simulated largest remnants immediately after the collision, but rather to those of a super-Earth with an assumed surface temperature of 1000 K and a core fraction $Z_\mathrm{Fe}$ after a long cooling period. Compared to the maximum stripping line (originally from \citet{marcus_minimum_2010}, and later corrected by \citet{reinhardt_forming_2022}), our maximum stripping line predicts slightly smaller radii for planets. Note that several super-Mercuries, such as Kepler-406\,b and Kepler-107\,c, fall just below the 80 km s$^{-1}$ head-on stripping line, suggesting that they could require a higher impact speed than 80 km s$^{-1}$ to form via impacts.
\footnotetext{The lines data can be found at \url{https://github.com/JingyaoDOU/Forming_S_Mercuries_2024}}

\subsection{Comparison with \citealt{denman_atmosphere_2020}}
We note that \citet{denman_atmosphere_2020} found a similar trend of $M_\mathrm{lr}$/$M_\mathrm{tot}$ as shown in their figure 4. \citet{denman_atmosphere_2020} found that, above an impact energy $Q_\mathrm{piv}$ at higher impact energies $Q_\mathrm{piv}$/$Q^*_\mathrm{RD}$, $M_\mathrm{lr}/M_\mathrm{tot}$ will have a steeper decreasing trend, and the loss efficiency of the atmosphere will decrease, which is due to the onset of mantle and core erosion. Their $Q_\mathrm{piv}$/$Q^{*}_\mathrm{RD}$ is close to our normalized impact energies where the first sharp change happens. The difference is that their $Q_\mathrm{piv}$/$Q^*_\mathrm{RD}$ is target mass dependent, and the break in $M_\mathrm{lr}$/$M_\mathrm{tot}$ slope occurs at slightly different $Q_\mathrm{R}$/$Q^{*}_\mathrm{RD}$ for different target mass. We infer that their changing of $Q_\mathrm{R}$/$Q^{*}_\mathrm{RD}$ is due to the different amount of atmosphere (ranging from 8-33 percent) in their three different target planets. For targets with low atmosphere mass ratio, the impact energy required to expel enough atmosphere for the core and mantle to start being lost is smaller compared to targets with higher atmosphere mass ratio. Combined results from their work and this work show that a universal law having a single trend may not be enough to describe all the collision processes that happen within multi-layered planets well. The universal law from \citet{stewart_velocity-dependent_2009, leinhardt_collisions_2012} works well for undifferentiated planetesimals, but when dealing with larger planets with two, three, or even four layers with oceans and atmospheres, due to the different shock impedances of different materials, the mass loss shows different trends.

\subsection{Polar head-on impact}

\begin{table*}
\centering
\caption{\label{tab:impact}Potential giant impact formation of an example super-Mercury with mass 5.0 M$_{\oplus}$ and 80\% core fraction. $\mathrm{M_{targ}}$ mass of initial pre-impact target in the unit of Earth-mass; $\gamma$ mass ratio between target and projectile; $V_\mathrm{imp}$ impact speed in km s$^{-1}$; $b$ impact parameter (zero indicates head-on equatorial impact if hit direction is -X, pole-on impact if hit direction is -Z); $P$ spin period of a pre-impact target in hours; $M_{lr}$ mass of the largest remnant after 200 hours simulation time; $Z_{Fe}$ iron element mass fraction in the largest remnant after a collision. Impacts are set up with the centre of mass of two impactors located at the origin with zero net velocity. For equatorial head-on impacts, impactors are moving toward each other along the X axis with zero Y and Z direction velocities, while for polar impacts, impactors are moving toward each other along Z axis with zero X and Y direction velocities. Hit direction -X and -Z means that the projectile is colliding with the target with the velocity pointing to the negative side of the X and Z axes, respectively. }
\begin{tabular}{ccccccccc}
\hline
Run &
$M_{\rm target}/M_{\oplus}$ &
$\gamma$ &
$V_{\rm imp}$ [km s$^{-1}$]&
$b$ &
$P$ [hr] &
$M_{\rm lr}$/$\mathrm{M_{\oplus}}$ &
$Z_{\rm Fe}$ &
hit direction\\
\hline
\noalign{\smallskip}
1 & 16.21 & 1.00 & 83.8 & 0 & 0 & 5.10 & 78.9\% & -X\\
\noalign{\smallskip}
\hline
\noalign{\smallskip}
2& 11.90 & 1.00 & 76.0 & 0 & 1.9 & 5.15 & 78.3\% & -Z\\
\noalign{\smallskip}
\hline
\end{tabular}
\end{table*}

Figure \ref{fig:mtar_mlr_zfe} illustrates that, according to the current scaling laws, a target mass of at least 10 M$_{\oplus}$ is typically required to generate a relatively dense remnant with mass greater than 5 M$_{\oplus}$ and core fraction greater than 60\%. To form a significantly denser planet, the required target mass would need to exceed 20 M$_{\oplus}$. However, \citet{crida_runaway_2017} suggests that planets with a mass above a critical value of 20 M$_{\oplus}$ must accrete gas and grow large enough to create a gap in the disk, in order to avoid being consumed by the star during fast Type-I migration. Given that planets with greater mass are more likely to undergo runaway accretion, it would be preferable to have targets with smaller mass as candidates for forming dense planets.

For high target mass impacts, head-on impacts still would be the most efficient way to strip off large amounts of mantle. To form a massive super-Mercury (e.g. 5.0 M$_{\oplus}$ with 80\% core fraction) via a giant impact, either requires an extremely energetic head-on impact resulting in catastrophic disruption of a target or some new impact configurations. If pre-impact spin is included, the effective gravitational potential is reduced and in extreme spinning cases \citep{lock_structure_2017,lock_origin_2018} the target is deformed making the core more exposed on the poles. To generate super-Mercuries with relatively low target masses and impact velocities compared to common head-on impacts, we propose an alternative impact configuration involving a fast-spinning target to increase the post-collision remnant iron mass fraction.

An erosive equal mass polar impact onto a fast-spinning proto-planet could potentially create a planet with a higher core fraction and larger remnant mass with lower impact velocity compared to common head-on giant impacts in the equatorial plane. A fast-spinning target with a spin period close to its instability rotational limit will deform greatly from a sphere to an oblate spheroid with mantle material concentrated at the equator. The core material is then closer to the surface in the polar regions. The core from an impactor can more easily penetrate through the thin mantle near the polar region of the target efficiently mixing with the target core. The impact shock generated from the polar impact will tend to kick more mantle material off than an impact onto a non-spinning target. 

For example, run 2 in Table \ref{tab:impact} (a pole-on impact with a fast spinning target, P =$ 1.9$ hr, mass = 11.9 M$_{\oplus}$ and an equal mass non-spinning projectile with an impact velocity of 76 km s$^{-1}$) generates a post-collision body with a mass of 5.15\,$\mathrm{M_{\oplus}}$ and core fraction $78.3\%$. The pole-on impact requires less total impact mass (23.8\,$\mathrm{M_{\oplus}}$) and lower impact speed (76 km s$^{-1}$) in contrast with the equatorial head-on impact which requires 32.4\,$\mathrm{M_{\oplus}}$ and an impact speed 83.8 km s$^{-1}$ (run 1 in Table \ref{tab:impact})  . 

Through stochastic accretion and migration stages, embryos and proto-planets can have a significant tilt with respect to their orbit. Also, rapidly rotating bodies are a common outcome of off-axis collisions, and average angular velocity after accretion or several giant impacts could be as high as the critical angular velocity \citep{kokubo_formation_2010,lock_structure_2017,nakajima_melting_2015}. Adopting a realistic accretion model, $N$-body simulations \citep{kokubo_formation_2010} show that spin obliquity could range from 0$^\circ$ to 180$^\circ$ following an isotropic distribution. Hence, the pole-on impact would not need a projectile embryo to have an orbit perpendicular to the orbital plane of the target. More detailed study of polar impacts in a wider parameter space will be carried out in future work.

\subsection{Vaporization vs kinetic kicking}

\begin{figure*}
\centering
\includegraphics[width=0.8\textwidth]{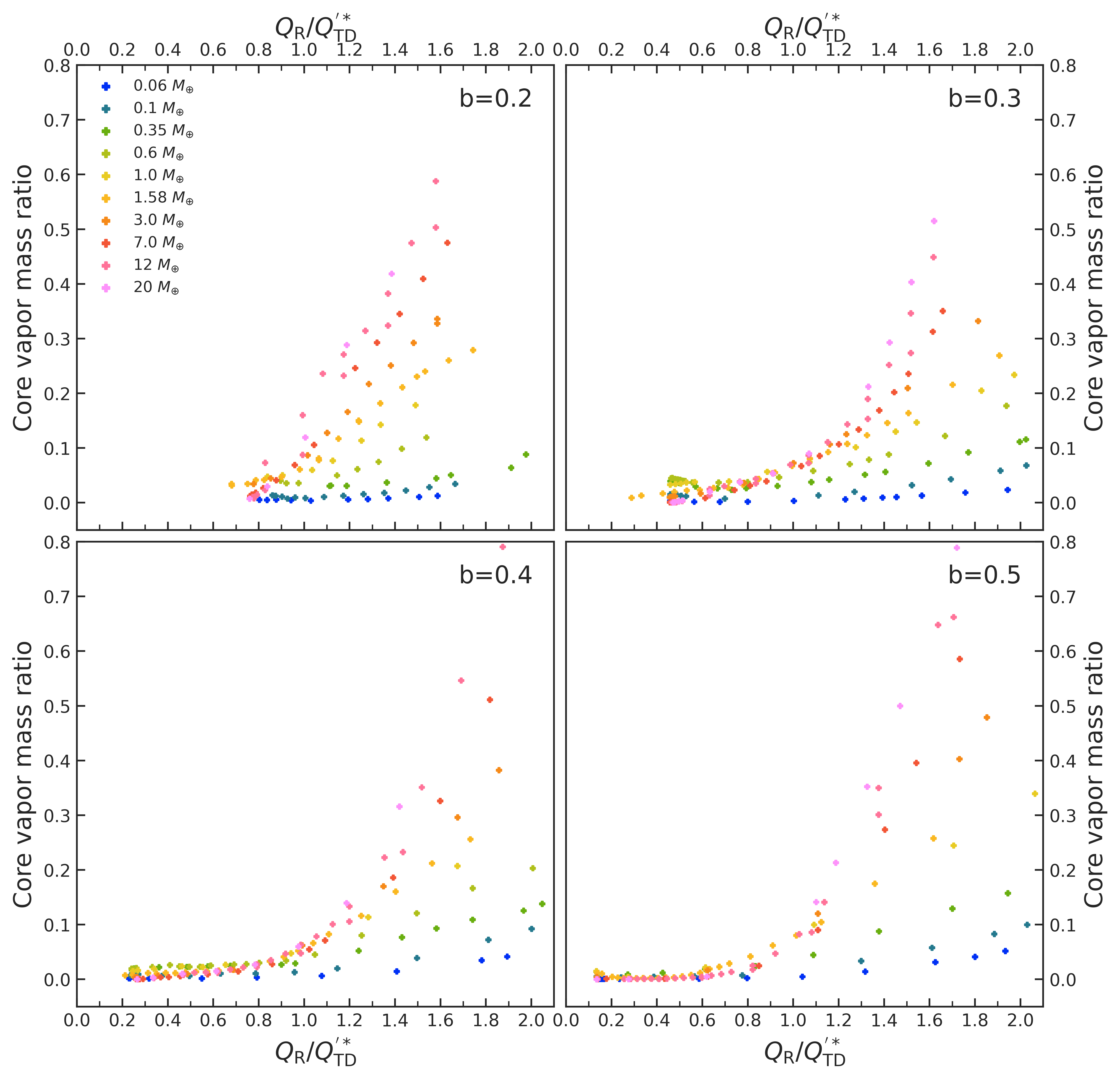}
    \caption{Fraction of vaporized core material of different target masses at various impact parameters. }
    \label{fig:ob_core_vap}
\end{figure*}

Transiting from head-on to oblique impacts, at low impact parameters, there will still be a significant amount of intersection between the target and impactor cores along their initial velocity vector. Therefore, core vaporization could still contribute to the mantle stripping process at relatively low impact angles. We examine core vaporisation in Figure \ref{fig:ob_core_vap} for impacts at different impact parameters. At $b=0.2$ and $b=0.3$, impacts with target mass $M_\mathrm{targ} =0.06$ M$_{\oplus}$ and 0.1 M$_{\oplus}$ have nearly zero core vaporization and therefore only kinetic kicking contributes to the mantle stripping process, leading to relatively lower mantle stripping efficiency compared to higher target mass impacts which have contributions from both kinetic kicking and core vaporization. At higher impact parameters ($b = 0.4$ and $b = 0.5$), low target mass impacts still have a low core vapor fraction. However, their iron mass fraction can be as high as high target mass impacts. This suggests that kinetic momentum transfer dominates the mantle stripping process at high impact angles. 

The iron mass fraction of the very small target mass impacts ($M_\mathrm{targ} =0.06$ and $M_\mathrm{targ} =0.1$) at $b=0.2$ and $b=0.3$ ( blue and green square and circle symbols in Figure \ref{fig:oblique}C respectively) is roughly 5\%-10\% lower than the higher target mass impacts. However, these low target mass impacts still generally have higher iron mass fraction than the corresponding head-on impacts which have maximum iron mass fraction around 50\% (see Figure \ref{fig:head-on}C). This also shows the limited role of core vaporization in oblique impacts. Impacts at $b=0.2$ represent the transition from head-on impacts to normal oblique impacts. Below this impact parameter, both kinetic momentum transfer and vaporization-induced ejection can both greatly affect the result of impacts. As the impact parameter increases, the intersection between the cores becomes smaller, and mantle stripping is gradually dominated by the kinetic ejection of mantle materials. For head-on impacts, the difference between mantle loss and core loss is around 30\% after the initial loss of the core, as shown in Figure \ref{fig:core_mantle_loss}. However, due to the misalignment nature of oblique impacts, the difference could be as high as 70\%. As a result, unlike vaporization dominated ejection of mantle for head-on impacts, mantle loss for oblique impacts is mainly dominated by kinetic kicking. Therefore, even low target mass impacts can form dense post-collision remnants.

At high impact parameters ($b=0.5$ and $b=0.7$) and high impact energy ($Q_\mathrm{R}$/$Q^{'*}_\mathrm{TD}>1.3$), the iron mass fraction of impacts with a target mass larger than 7 M$_{\oplus}$ is surprisingly lower compared to smaller target mass impacts (Figure \ref{fig:oblique}D). Due to the extremely high impact energies, the core and mantle materials in these impacts with high target masses and high impact parameters are mostly in a supercritical state after the collision. The difference in iron mass fraction at these high impact parameters may be due to the different phase transition processes experienced by the core and mantle materials during normal and extremely high energy impacts. For these high target mass impacts, a greater amount of kinetic energy is dissipated due to the phase transition, which might in turn lead to lower mantle stripping efficiency.  Furthermore, since these impacts occurred at extremely high impact velocities ($V_\mathrm{i}>200$ km s$^{-1}$) and energies, the EoS tables we used may not be sufficient to model the state of the material under these extreme conditions. We plan to investigate this further in our future work.

\subsection{Caveats}

Although the revised M-ANEOS equations of state \citep{stewart_equation_2020,stewart_equation_2019} can better model phase transitions, we extended these EoS tables to higher maximum densities (100\,$\mathrm{g\,cm}^{-3}$ for forsterite and 200 \,$\mathrm{g\,cm}^{-3}$ for iron) which exceed the phase space that the tables were originally designed to deal with into a region with no experimental data to confirm material properties.
This will mainly affect the high target mass impacts at high impact energies. The mass and iron mass fraction for high target mass impacts shown in Figure \ref{fig:head-on}B and D do not show significantly different behavior at high impact energies compared to relatively low target mass impacts, which suggests any error introduced by extending the EoS tables is negligible. 

All simulations in this work used the same target and impactor for each impact. Therefore, the scaling laws derived from these simulations apply conservatively only to equal-mass impacts. To better understand the impact conditions that can form dense planets, we tested oblique impacts with impact parameters up to $b=0.7$, so all the fitted scaling laws are valid below $b=0.7$. For impact parameters above $b=0.7$, the probability of having energetic erosive hit-and-run events is very low. Additionally, the impact velocities required to form a post-collision remnant with an iron mass fraction larger than 60\% could easily exceed ten times the mutual escape velocities.

As we expect lower mass projectiles to cause less erosion for a given impact energy, equal-mass impacts provide a useful `lower limit' and our study can provide useful reference to constrain the formation conditions of super-Mercuries. Although using only equal-mass impacts narrows the parameter space, the data is consistent and allows for more accurate fitting of scaling laws. %Compared to impacts with $\gamma$ less than one, equal-mass impacts represent the minimum impact energies needed to form a super-Mercury for a specific target mass.

All impact simulations in this work only represent the collision outcome immediately after giant impacts, within about 30 hours. This does not account for cooling or the effect of re-accretion of the ejecta after a long period of time. \citet{benz_origin_2007} reported that up to approximately 40\% of ejected materials will accumulate back to Mercury after several million years. However, the amount of re-accumulation depends on the configuration of an impact and the specific structure of the planetary system. In addition,  \citet{spalding_solar_2020} showed that the primordial solar wind could provide sufficient drag on ejected debris to remove them from Mercury-crossing trajectories before re-accreting back to the planet's surface. This suggests that the giant impact hypothesis remains a viable pathway toward the formation of dense planets, particularly those close to young stars. Therefore, the results derived from this work give an upper limit of the core fraction and a lower limit of the mass of super-Mercuries formed by giant impacts.

\subsection{Dilemma for the formation of super-Mercuries}
So far, more than 5000 exoplanets have been discovered but less than ten planets are confirmed to be candidate super-Mercuries. Head-on super-Earth impacts theoretically are the most efficient mantle stripping method, however, compared to oblique impacts, they are very rare. Thus, impact generated super-Mercuries would most likely be formed by one or a sequence of oblique giant impacts. Oblique impacts are more likely happened around 45$^{\circ}$ or $b=0.7$, with probability decreasing with decrease of impact angle. However, at high impact parameter as $b=0.7$,  the mantle stripping efficiency is extremely low as shown in the Figure \ref{fig:oblique} and \ref{fig:v_z_m}. At $b=0.7$, the required impact velocities to form a super-Mercury with 60\% core even above 1 M$_{\oplus}$ is unrealistically high. 
Balancing the low probability of impacts at low impact parameters and low mantle stripping efficiency at high impact parameters, we conclude that impact generated super-Mercuries would more likely occur with impacts between $b=0.3$ and $b=0.5$ (17$^{\circ}$ to 30$^{\circ}$).

It is also worth noting that, previous $N$-body simulations normally have an inner semi-major axis cutoff between 0.1 and 0.5 au, while several potential super-Mercuries have even smaller orbits around star, e.g. Kepler-107\,c has orbit radius 0.06 au and GJ 367\,b has orbit around 0.007 au. The conditions of impacts that occur at such short orbital distances needs further study in order to determine the likelihood of super-Mercury forming collisions.

\section{CONCLUSIONS}
\label{sec:conlcusion}
In this work, we present over one thousand giant impact smoothed particle hydrodynamics simulations, ranging from head-on to oblique impacts. All simulations feature equal-mass impacts, with target mass spanning a wide range from 0.06 M$_{\oplus}$ to 20 M$_{\oplus}$. We find that the vaporization of core materials during collisions can greatly affect the outcome of giant impacts. We propose that the stripping of mantle materials is mainly controlled by two mechanisms: 1) core vaporization-enhanced stripping and 2) kinetic momentum transfer.
For target planets with a mass less than 1 M$_{\oplus}$, head-on impacts fail to form remnants with an iron mass fraction larger than 50\%, due to the lack of core vaporization. For small target mass impacts, we find that the largest remnant tends to fall into a new fragmentation regime, splitting into several smaller pieces at a small $Q_\mathrm{R}$/$Q^{*}_\mathrm{RD}$ value around 1.15, which corresponds to much smaller impact energies compared to those predicted by previous studies. We plan to further investigate the reasons causing fragmentation in our follow-up paper.
For target masses above 2 M$_{\oplus}$, the mass and iron mass fraction show a clear trend, and we fit a new scaling law for these impacts.

Oblique impacts do not show significant differences between the low and high target mass regimes.  We fit mass and iron mass fraction scaling laws for oblique impacts and propose the new target mass dependent hit-and-run velocity criteria. Based on our head-on data, we fit a more accurate free parameter $c^*$, 2.926, used to define the catastrophic disruption criteria for two-layered super-Earth impacts. We also propose a further angle correction for target catastrophic disruption criteria. Combining the catastrophic disruption criteria and the new scaling laws, one can easily predict the outcome both for head-on and oblique impacts. Based on our new scaling laws, we describe the necessary impact conditions required to form planets with various mass and iron mass fraction. Combining these with interior structure models for super-Earths, we update the `maximum mantle stripping line' at different impact velocities and angles. Additionally, we provide a prescription on how to use these scaling laws in the Appendix \ref{sec:prescription}.

We demonstrate that for oblique impacts, the efficiency of mantle stripping decreases significantly as the impact angle increases. At the most common impact angle of 45$^\circ$, an exceedingly high impact velocity is required to create a relatively dense planet. This presents a challenge in the formation of super-Mercuries through a single giant impact. However, the majority of currently observed dense planets have extremely short orbital periods and consequently high orbital velocities. This mitigates the restrictions on the impact conditions and makes the scenario of impact formation plausible.

These simulations provide important insights into the formation and evolution of dense planets. Future studies could build on our work by exploring a wider range of parameters and conditions, including different impactor-to-target mass ratio, the effect of spin, different chemical composition, different stellar type, and long-term evolution of the post-collision remnants.

Our work provides a useful framework for interpreting observational data of exoplanets and understanding the formation history of planetary systems. With more planets discovered in the future, we hope we can better understand the formation of our own Solar System and the prevalence and diversity of dense rocky exoplanets throughout the galaxy.

\section*{Acknowledgements}

We thank the anonymous reviewer for the constructive comments which have improved the quality of this manuscript. J. D. acknowledges funding support from the Chinese Scholarship Council (No. 202008610218). PJC and ZML acknowledge financial support from STFC (grant number: ST/V000454/1). The giant impact simulations were carried out using the computational facilities of the Advanced Computing Research Centre, University of Bristol – \url{http://www.bristol.ac.uk/acrc/} and Isambard 2 UK National Tier-2 HPC Service (http://gw4.ac.uk/isambard/) operated by GW4 and the UK Met Office, and funded by EPSRC (EP/T022078/1).

%%%%%%%%%%%%%%%%%%%%%%%%%%%%%%%%%%%%%%%%%%%%%%%%%%
\section*{Data Availability}

Full simulation output is available from the authors on reasonable request.
%The inclusion of a Data Availability Statement is a requirement for articles published in MNRAS. Data Availability Statements provide a standardised format for readers to understand the availability of data underlying the research results described in the article. The statement may refer to original data generated in the course of the study or to third-party data analysed in the article. The statement should describe and provide means of access, where possible, by linking to the data or providing the required accession numbers for the relevant databases or DOIs.



\bibliographystyle{mnras}
%\bibliography{ref} % if your bibtex file is called example.bib
%%%%%%%%%%%%%%%%%%%%%%%%%%%%%%%%%%%%%%%%%%%%%%%%%%%%%%%%%%%%%

%%%%%%%%%%%%%%%%%%%%%%%%%%%%%%%%%%%%%%%%%%%%%%%%%%%%%%%%%%%%%

%\begin{thebibliography}{99}
%\end{thebibliography}

%%%%%%%%%%%%%%%%%%%%%%%%%%%%%%%%%%%%%%%%%%%%%%%%%%

%%%%%%%%%%%%%%%%% APPENDICES %%%%%%%%%%%%%%%%%%%%%

\appendix

\section{Fitting results}
Below we show the fitted scaling laws for head-on impacts and oblique impacts.

\begin{figure}
    \includegraphics[width=0.9\columnwidth]{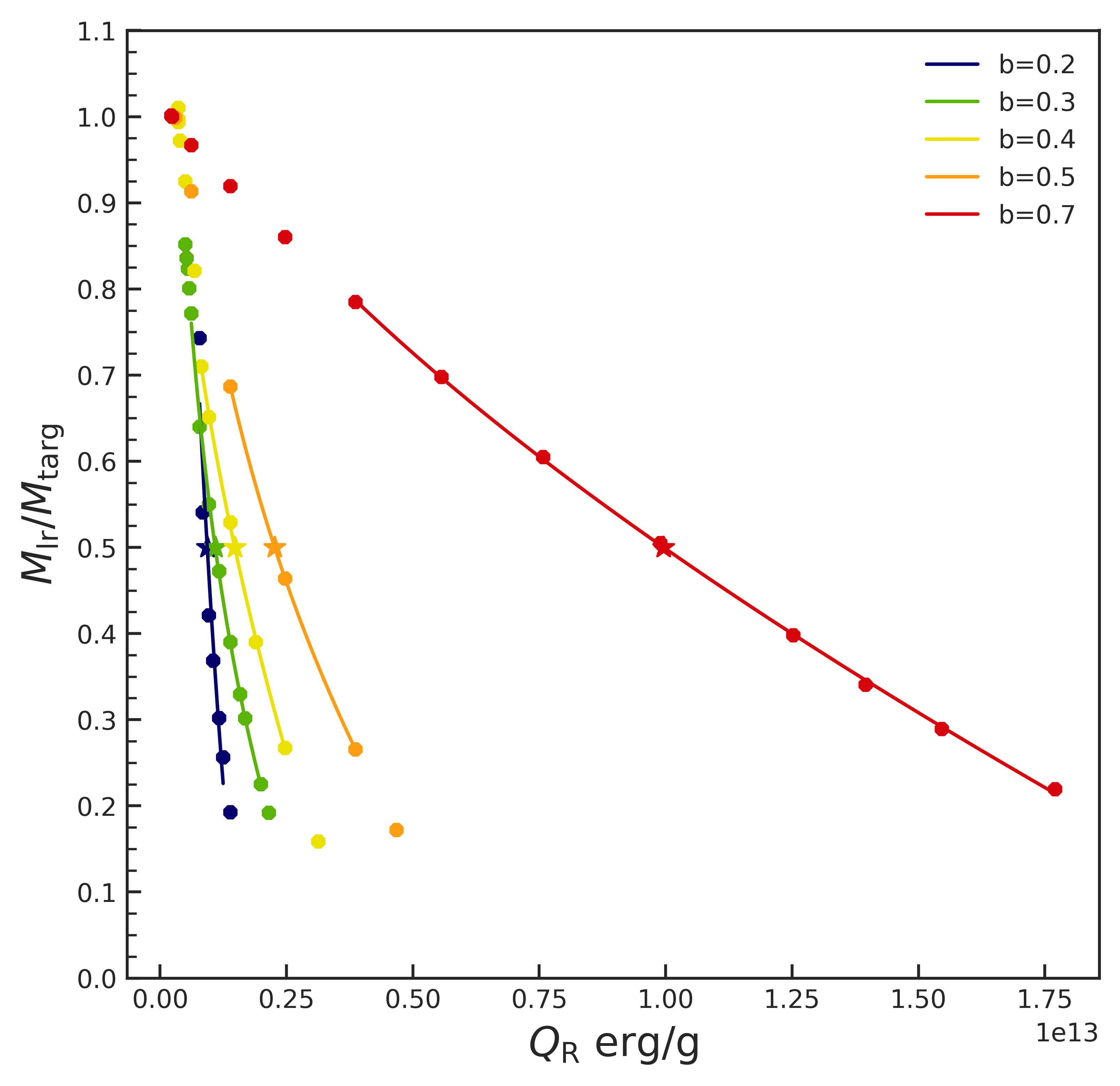}
    \caption{Demonstration of the fitting process for equation \ref{eq:QTD} to determine the target catastrophic disruption criterion $Q^{'*}_\mathrm{TD}$ for the target with mass 1\,M$_{\oplus}$ at various impact parameters. The star symbols represent the fitted $Q_\mathrm{R}$ when $M_\mathrm{lr}$/$M_\mathrm{targ}$ is 0.5 at different impact parameters.}
    \label{fig:m1d0_fitting}
\end{figure}

\begin{figure}
    \includegraphics[width=0.9\columnwidth]{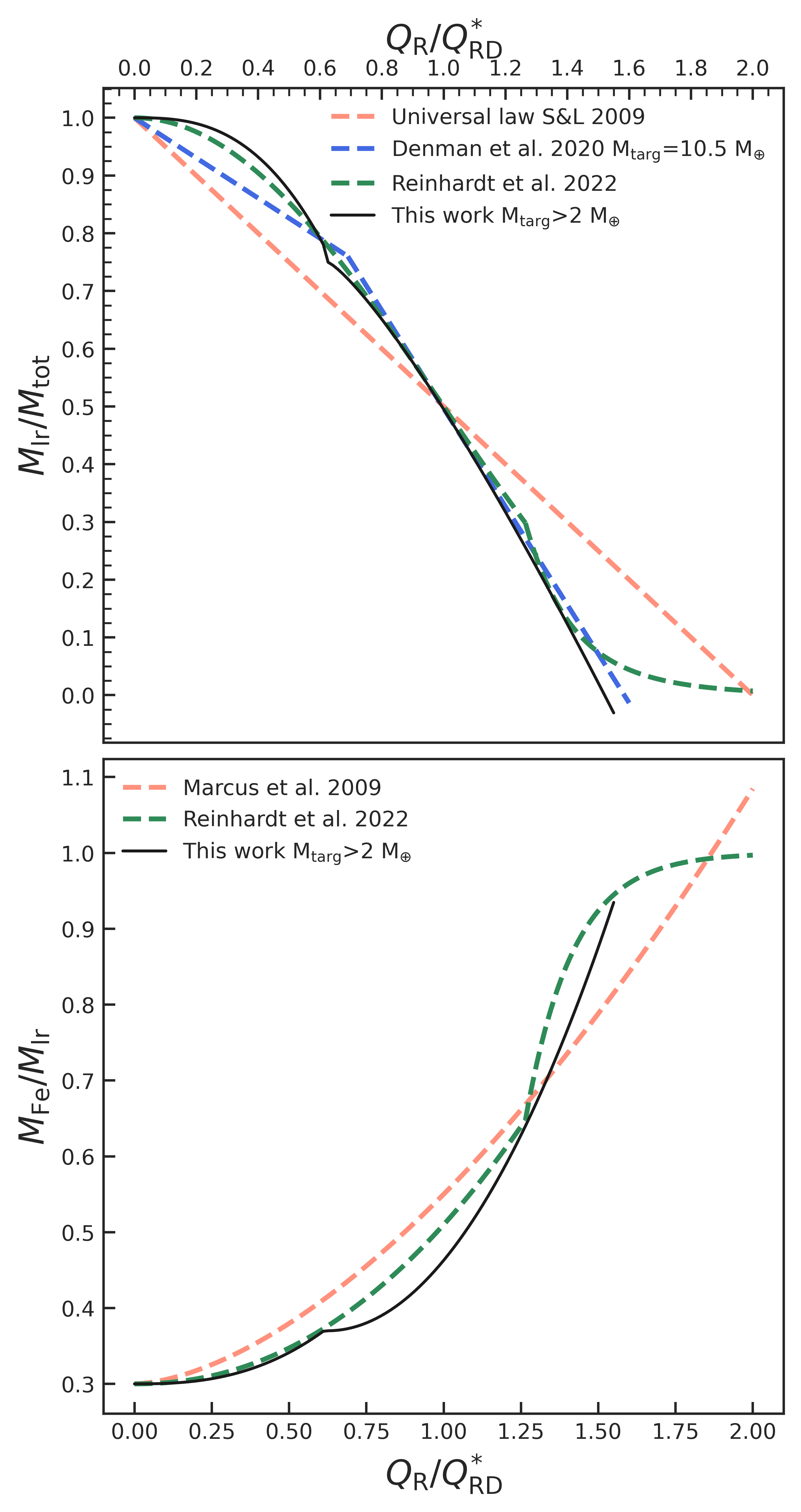}
    \caption{Comparison of scaling laws of mass (top panel) and iron mass fraction (bottom panel) of the largest post-collision remnant. In the top panel, the pink dashed line shows the universal law first derived in \citet{stewart_velocity-dependent_2009}. The blue dashed line shows the fitting result from \citet{denman_atmosphere_2020} at target mass around 10.5 M$_{\oplus}$. The green dashed lines shows the scaling laws from \citet{reinhardt_forming_2022}. In the bottom panel, the pink dashed line shows the iron mass fraction scaling law from \citet{marcus_collisional_2009} with the initial iron mass fraction shifted to 30\%.}
    \label{fig:head_on laws compare}
\end{figure}

\section{Other supplementary plots}

\begin{figure}
\includegraphics[width=0.9\columnwidth]{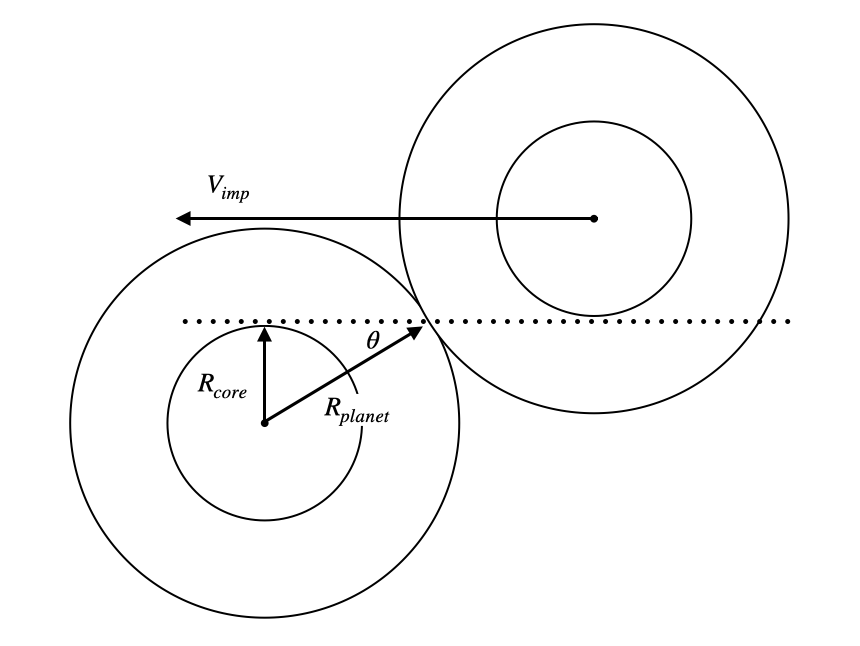}
    \caption{Schematic of the collision geometry. $\theta$ is the impact angle and the impact parameter is $b=\sin{\theta}$.}
    \label{fig:impact_geo}
\end{figure}

\section{Prescription to predict mass and iron mass fraction}
\label{sec:prescription}

Below, we describe the prescription that can be used to predict the mass and iron mass fraction of the largest remnant based on the scaling laws we derived in this work. The prescription should only apply to equal-mass single giant impacts with a target mass above 0.01 M$_{\oplus}$. The prescription for an impact at impact velocity $v_\mathrm{imp}$ and impact parameter $b$ is as follows:
\\
\\
\noindent(i) For a target and an impactor planet with masses and radii $M_\mathrm{targ}$, $M_\mathrm{imp}$, $R_\mathrm{targ}$ and $R_\mathrm{imp}$, first calculate the mutual escape velocity of the system:
\begin{equation}
    V_\mathrm{esc}=\sqrt{\frac{2G(M_\mathrm{targ}+M_\mathrm{imp})}{R_\mathrm{targ}+R_\mathrm{imp}}}.
	\label{eq:ves}
\end{equation}

\noindent(ii) Next, calculate the specific impact energy:
\begin{equation}
    Q_\mathrm{R}=0.5\mu\frac{V^2_\mathrm{i}}{M_\mathrm{tot}},
	\label{eq:QR}
\end{equation}
where $\mu=M_\mathrm{targ}M_\mathrm{imp}/M_\mathrm{tot}$.
\\
\\
\noindent(iii.a) \textbf{Head-on impact} ($b = 0$) Calculate catastrophic disruption criteria for head-on impacts:
\begin{equation}
    Q^{*}_{\mathrm{RD},\gamma=1} = c^{*}\frac{4}{5}\pi\rho_{1}GR^2_\mathrm{C1},
    \label{eq:c-star} 
\end{equation}
where $c^*$ equals 2.926, $\rho_1$ = 1000 kg m$^{-3}$, and $R_\mathrm{C1}$ = $(\frac{3 M_\mathrm{{tot}}}{4\pi\rho_1})^{\frac{1}{3}}$.
\\
\\
\noindent(1) If a target mass is less than 2 M$_{\oplus}$, using equation \ref{eq:head-on:mlr_lm} and equation \ref{eq:head-on:zfe_l} to predict the mass and iron mass fraction respectively. If $Q_\mathrm{R}$/$Q^*_\mathrm{RD}$ is larger than 1.2,  label the impact as fragmentation, indicating that no main remnant is left in the system.
\\
\\
\noindent(2) If a target mass is larger than 2 M$_{\oplus}$, using equation \ref{eq:head-on:mlr_hm} and equation \ref{eq:head-on:zfe_h} to predict the mass and iron mass fraction respectively.
\\
\\
\noindent(iii.b) \textbf{Oblique impact} ($b\geq0.2$) Calculate the target catastrophic disruption criteria:
\begin{equation}
    Q^{'*}_\mathrm{TD} = \left(0.175 \left(\frac{1}{1-b} \right)^{3.3}+0.635\right)c^* \frac{4}{5}\pi\rho_{1}GR^2_\mathrm{C1},
    \label{eq:Q_td}
\end{equation}
where $b$ is impact parameter and $c^*$ equals 2.926.
\\
\\
\noindent(1) Find the impact velocity where hit-and-run starts to happen: 
\begin{equation}
    V_\mathrm{HnR} = \left(\Gamma(1-b)^{3.6}+1.2\right)V_\mathrm{esc},  
    \label{eq:v_hnr}
\end{equation}
where $\Gamma = -0.3\log_{10} M_{\mathrm{targ},\oplus} + 2.18$, and $M_{\mathrm{targ},\oplus}$ is the mass of target planet in Earth masses.
\\
\\
\noindent(2) If V$_\mathrm{i}$ is less than V$_\mathrm{HnR}$, then an impact should be treated as perfect merging. 
\\
\\
\noindent(3) Else, if V$_\mathrm{i}$ is larger than V$_\mathrm{HnR}$ and impact parameter $b$ >= 0.3, equation \ref{eq:oblique-mlr} and equation \ref{eq:oblique-zfe} should used to predict the mass and iron mass fraction. The parameters $\alpha_{M,b}$, $\alpha_{Fe,b}$, and $\beta_{Fe,b}$ in equations \ref{eq:oblique-mlr} and \ref{eq:oblique-zfe} can be calculated by the equations \ref{eq:amlr}, \ref{eq:afe}, and \ref{eq:bfe}.
\\
\\
\noindent(4) If $b$ is between 0.2 and 0.3, equation \ref{eq:oblique_fit_b0d2} and equation \ref{eq:oblique-zfe} (with $\alpha_{Fe,0.2}=0.199$, and $\beta_{Fe,0.2}=1.904$) should be used to approximate the mass and iron mass fraction of the largest remnant.
\\
\\
\noindent(5) If $b$ is between 0.0 and 0.2, then an impact should be approximate as head-on impact and using prescription described above for head-on impact.

%%%%%%%%%%%%%%%%%%%%%%%%%%%%%%%%%%%%%%%%%%%%%%%%%%

% Don't change these lines
\bsp	% typesetting comment
\label{lastpage}
\end{document}